\documentclass[
 reprint,
superscriptaddress,
 amsmath,amssymb,
 aps,
]{revtex4-2}
\usepackage{mathtools}
\usepackage{xcolor}
\usepackage{graphicx}
\usepackage{dcolumn}
\usepackage{bm}
\usepackage{siunitx}
\usepackage[utf8]{inputenc}
\usepackage[T1]{fontenc}

\makeatletter
\renewcommand*{\fnum@figure}{{\normalfont\bfseries \figurename~\thefigure}}
\renewcommand*{\@caption@fignum@sep}{\textbf{\usepackage{.} }}
\makeatother
\usepackage{amsmath}
\usepackage{setspace}
\usepackage {algpseudocode}
\usepackage{lipsum}
\usepackage{xurl}
\makeatletter
\renewcommand*{\fnum@figure}{{\normalfont\bfseries \figurename~\thefigure}}
\renewcommand*{\@caption@fignum@sep}{\textbf{. }}
\makeatother
\renewcommand{\thefigure}{\arabic{figure}}
\usepackage{hyperref}

\begin{document}

\preprint{APS/123-QED}
\raggedbottom
\title{Intrinsically Design-Rule-Compliant Nanophotonic Inverse Design via Learned Generative Manifolds}

\author{Bahrem~Serhat~Danis}
\affiliation{\mbox{Dept. of Electrical and Electronics Engineering, Koç University, Istanbul, 34450, Turkey}}

\author{Demet~Baldan~Desdemir}
\affiliation{\mbox{Dept. of Electrical and Electronics Engineering, Koç University, Istanbul, 34450, Turkey}}

\author{Enes~Akcakoca}
\affiliation{\mbox{Dept. of Electrical and Electronics Engineering, Koç University, Istanbul, 34450, Turkey}}

\author{Zeynep~Ipek~Yanmaz}
\affiliation{\mbox{Dept. of Electrical and Electronics Engineering, Koç University, Istanbul, 34450, Turkey}}

\author{Gulzade~Polat}
\affiliation{\mbox{Dept. of Electrical and Electronics Engineering, Koç University, Istanbul, 34450, Turkey}}

\author{Ahmet~Onur~Dasdemir}
\affiliation{\mbox{Dept. of Electrical and Electronics Engineering, Koç University, Istanbul, 34450, Turkey}}

\author{Aytug~Aydogan}
\affiliation{\mbox{KU Leuven, Dept. of Physics and Astronomy, B-3000 Leuven, Belgium}}

\author{Abdullah~Magden}
\affiliation{\mbox{Dept. of Mathematics, Faculty of Engineering and Natural Sciences, Bursa Technical University, Bursa, 16310, Turkey}}

\author{Emir~Salih~Magden}
\email{Corresponding author: esmagden@ku.edu.tr}
\affiliation{\mbox{Dept. of Electrical and Electronics Engineering, Koç University, Istanbul, 34450, Turkey}}
\affiliation{\mbox{KUIS AI, Koç University, Istanbul, 34450, Turkey}}


\begin{abstract}
Inverse design has enabled the systematic design of ultra-compact and high-performance nanophotonic components. Yet enforcing design-rule compliance throughout the entire optimization trajectory, without platform-specific regularization scheduling, remains an open problem. Without explicit fabrication constraints, optimized geometries can violate minimum feature size constraints, producing structures incompatible with lithography processes. Established projection-filter and morphological constraint methods enforce compliance within topology optimization, but rely on scheduled regularization with process-dependent design rules. This motivates parameterization approaches that encode fabrication constraints as intrinsic properties of the design space. Here, we demonstrate intrinsically design-rule-compliant silicon photonic inverse design through a deep generative reparameterization that confines optimization to a learned manifold of fabrication-compatible geometries, reducing computational cost by 5-fold over unconstrained pixel-based methods. We validate this approach across representative silicon photonic devices including broadband power splitters, spectral duplexers, and mode converters operating across the 1,500--1,600 nm band, for both electron-beam lithography and photolithography platforms. Across all devices, the manifold-based formulation achieves competitive performance metrics, while ensuring design-rule compliance throughout the entire optimization trajectory. By treating fabrication constraints as a fundamental property of the design representation rather than an external penalty, this work establishes a direct pathway toward automated, platform-agnostic, design-rule-compliant nanophotonic design pipelines.
\end{abstract}

\maketitle


\section{Introduction}

Integrated photonic components are indispensable for high-speed communications~\cite{Shekhar2024, Li2023}, sensing~\cite{Vollmer2008}, and computation~\cite{Bogaerts2020, ChenX2023}, as they combine low propagation losses with wide bandwidths~\cite{Cheben2018, ZhangL2022, NajjarAmiri2024, VitAycanPolHandling2025, VitAycanFabTol2025, GorguluKazimDispersion2025} in compact footprints~\cite{Chang2022, Grotevent2023, MehmetCanOktayEME2024, MehmetCanOktayEMEConference2025, AysemineConference2024, AhmetOnurDasdemirConference2020}. Achieving such performance, however, requires design methodologies that go beyond fundamental understanding of light propagation~\cite{Tran2022}, mode coupling dynamics~\cite{Hinney2024}, or typical designer intuition~\cite{PeterWiecha2021}. In this domain, inverse design has emerged as a powerful paradigm for realizing photonic devices with superior performance, as many of these geometries are prohibitively complex to create manually~\cite{Molesky2018, MacLellanBenjamin2024, RasmusChristiansen2021, Hughes2018, KaiyuanWang2020}. By leveraging algorithmic design procedures, many different classes of devices, including wavelength-dependent and polarization-sensitive optical signal processors~\cite{AlexanderPiggott2015, MaoqingGuo2025}, high-speed modulators~\cite{WeibaoHe2021, Lin2017}, and compact multimode grating couplers~\cite{Yang2022}, have already been demonstrated. In these design approaches, a target device response is first specified, and an algorithm iteratively refines a set of geometrical parameters to improve a quantitative figure of merit through gradient-based optimization. The material topology or shape distribution within the design region is optimized by iteratively updating parameters according to these gradients, ultimately converging toward geometries that achieve the desired performance objectives~\cite{ChristopherM2013, Jensen2011, HootenSean2025}. Even though inverse design of photonic structures has produced impressive results, the underlying optimization problem is inherently ill-posed~\cite{SchwabJohannes2019}. For a given target optical operation, there typically exists not a single unique geometry but a vast family of structures that all yield acceptable results in terms of extinction ratio, insertion loss, or related performance metrics. Moreover, designers typically seek a solution that performs acceptably well rather than the absolute global optimum, further broadening the solution space~\cite{Zandehshahvar2022, XuXiaopeng2023, WongJoshua2026}. This reflects a broad, high-dimensional set of geometries that effectively yield indistinguishable optical responses, or perform equally well within a given application space. This non-uniqueness is typically handled in inverse problems by using additional regularizations or constraints \cite{SchwabJohannes2019, GertlerShai2025} that guide convergence towards solutions accommodating additional design requirements beyond optical performance alone.

A particularly critical one of these requirements is compatibility with fabrication processes. Despite the abundance of optically equivalent permittivity distributions, practical use and scalability are only possible for a small subset of these geometries that are also compliant with fabrication rules. While the large non-unique solution space points to the existence of solutions that are fabrication-compatible, optimization methods face difficulties efficiently navigating toward devices that also satisfy fabrication constraints. Specifically, layouts of fabrication-compatible devices must satisfy design rules such as minimum feature size, curvature limits, enclosed-area requirements, and angle-of-curvature conditions. If these rules are not enforced during the design stage, the resulting geometries risk being fabrication-incompatible, leading to partial pattern transfer, degraded performance, or complete device failure~\cite{PiggottAlexander2017, PiggottAlexander2020}. A rigorous class of fabrication-constrained inverse design methods has been 
developed within the topology optimization community. Systematic projection and 
density filters imposing minimum feature sizes and spacing constraints were used 
\cite{WangFengwen2022, LazarovBoyan2016}, building on the morphological and density-based 
optimization frameworks \cite{Jensen2011, RasmusChristiansen2021}. Building on this 
foundation, semiconductor foundry design rules including minimum feature size, 
minimum spacing, and enclosed area constraints can be directly embedded within 
photonic topology optimization with hyperparameter scheduling to achieve 
high-performance, fabrication-compliant devices \cite{HammondAlec2021}. Due to the sensitivity of these hyperparameter choices, it is also possible that optimizations may require several restarts or fine-tuning of the optimization schedule depending on device functionality or fabrication platforms, potentially limiting the effectiveness of design frameworks and scalability~\cite{HammondAlec2021, VercruysseDries2019, SeoDongjin2026, ChenYuchen2023, HiesenerJacob2025}. Other strategies construct devices entirely from smaller, rule-compliant building blocks~\cite{KaiyuanWang2020, SchubertMartin2022, TangYingheng2020} or phase-projection-based generators~\cite{ChenHao2024}. While this guarantees fabrication compatibility by design, the restricted design space often forces a trade-off, sacrificing either high performance or overall footprint compactness for the desired operation.

The vast solution space points to the existence of fabrication-compatible designs within. However, currently available optimization approaches require external penalty terms and careful hyperparameter scheduling to reach them. These two factors motivate a fundamentally different design paradigm. In this paper, we introduce a framework for nanophotonic inverse design that utilizes a generative reparameterization to directly parameterize fabrication-compatible devices as a low-dimensional manifold of design rule check (DRC)--compliant geometries within the much larger set of possible solutions. By reparameterizing the design search, the ill-posed optimization problem is transformed into a tractable, intrinsically fabrication-compliant framework. Our generator network is trained with fabrication-compatibility metrics that enforce minimum feature size and discrete topology while ensuring compatibility with standard CMOS design rules. We demonstrate this framework across a variety of devices including power splitters, wavelength multiplexers/demultiplexers, as well as mode converters, and show that optimization converges in significantly fewer iterations while maintaining fabrication compliance throughout. Furthermore, by simply adjusting the spatial resolution of the generator layers, we demonstrate that the same approach can be applied to different fabrication platforms, from electron-beam lithography (EBL, $60\,\text{nm}$ minimum feature size) to photolithography (PL, $150\,\text{nm}$ minimum feature size). This integration yields high-performance, fabrication-compatible devices while significantly reducing optimization time, offering a scalable pathway toward the next generation of design-rule-compliant integrated photonic technologies.

\section{Manifold Reparameterization for Intrinsically Fabrication-Compliant Inverse Design}

\begin{figure}[ht!]
    \centering
    \includegraphics[width=\columnwidth]{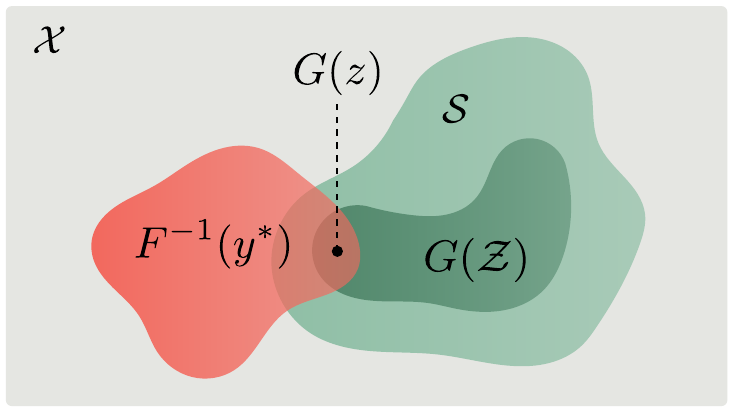}
    \caption{Conceptual illustration of search domains and solution sets in inverse design. The ambient design space is denoted by $\mathcal{X}$. The set $F^{-1}(y^{*})$ corresponds to all permittivity distributions that achieve the desired target response $y^{*}$. The subset $\mathcal{S} \subset \mathcal{X}$ represents the space of fabrication-compatible permittivity distributions. DRC-compliant solutions therefore lie in the intersection of $F^{-1}(y^{*}) \cap \mathcal{S}$. The generative manifold $G(\mathcal{Z})$, parameterized by the latent space $\mathcal{Z}$, is embedded within the subset $\mathcal{S}$, ensuring fabrication compatibility by construction. $G(z)$ denotes one feasible solution within the manifold.}
    \label{fig:figure1}
\end{figure}

In conventional pixel-based inverse design, achieving a high-performance device involves continuously updating the design parameters that describe the permittivity distribution of the core and cladding regions. In practice, the design region is discretized into a grid of pixels, and the value of each pixel’s permittivity is iteratively adjusted during device search until a distribution yielding desirable device performance is obtained. Each vector $x$ representing a pixel-based permittivity distribution in the search space. The set of all possible vectors therefore forms the design space $\mathcal{X}$, as illustrated in Fig.~\ref{fig:figure1}. Since conventional inverse design explores the permittivity distribution explicitly via a direct search over pixel-based solutions, this space $\mathcal{X}$ functions as an ambient search domain. During device optimization, the proximity of the current device response to the target response is evaluated through electromagnetic solvers. These solvers take the permittivity distribution as input and produce the corresponding device response, formally described by a response function $F(x)$. The goal is to identify a permittivity distribution that yields the desired target response $y^*$. The device search systematically explores candidate distributions $x \in \mathcal{X}$ to find an appropriate solution that satisfies $F(x) = y^*$. However, since different permittivity distributions can produce similar or even identical responses, this inverse problem is inherently ill-posed as many non-unique solutions exist for a given $y^*$. The set of all these solutions is a subspace of $\mathcal{X}$ denoted by $F^{-1}(y^*)$ in Fig.~\ref{fig:figure1}. As elaborated below, this non-uniqueness, and the resulting richness of the set $F^{-1}(y^*)$, is precisely what points to the existence of the intersection $F^{-1}(y^*) \cap \mathcal{S}$ and is therefore the main leverage employed in our manifold-based approach. In addition to finding appropriate solutions, the optimization process also incorporates a regularization term $R(x)$ governing the transformations applied to the permittivity distribution such as the scheduling of spatial averaging, filtering, and projection operations. Physically, $R(x)$ enforces design-rule compliance including minimum feature size, minimum enclosed area, and curvature constraints, while guiding the distribution toward a well-defined material boundary form suitable for fabrication. This forces the optimization to converge to a solution within $\mathcal{S}$, the set of all fabrication-compatible devices where $\mathcal{S} \subset \mathcal{X}$. Mathematically, this inverse problem can thus be expressed as

\begin{equation}
\arg \min_{x \in \mathcal{X}} \Big\{ \| F(x) - y^{*} \|^2 + R(x) \Big\}.
\label{eq:regularization_inv_problem}
\end{equation}
Because the inverse mapping $F^{-1}(y^{*})$ is inherently ill-posed, regularizers 
of this form have become popular for achieving fabrication compatibility. However, 
the ill-posedness of $F^{-1}(y^*)$ is generally not a problem that prevents 
convergence to acceptable solutions~\cite{WangFengwen2022, LazarovBoyan2016, HammondAlec2021}. In fact, the richness of the non-unique 
solution space is precisely what makes the intersection $F^{-1}(y^*) \cap 
\mathcal{S}$ accessible by search algorithms. The difficulty arises because 
conventional methods operate in the ambient space $\mathcal{X}$ without an 
intrinsic mechanism to restrict the search to the DRC-compliant subset 
$\mathcal{S}$. While regularizers are indispensable for enforcing these 
constraints, they often complicate the optimization landscape due to their 
dependence on scheduling strategies and hyperparameter selection. The resulting 
increase in iterations required for convergence motivates alternative 
parameterization strategies that encode fabrication constraints structurally, 
effectively eliminating the need for explicit regularization terms.

Instead of conducting the search directly in the ambient design space $\mathcal{X}$, we introduce a fundamentally different methodology in which the inverse design problem is reparameterized on a low-dimensional manifold $G(\mathcal{Z})$, where each $G(z) \in G(\mathcal{Z})$ represents a single fabrication-compatible device geometry in the latent space $\mathcal{Z}$. Consequently, all solutions in this low-dimensional manifold $G(\mathcal{Z})$ also exist in $\mathcal{S}$, effectively ensuring that $G(\mathcal{Z}) \subset S$. Within $G(\mathcal{Z})$, as optimization is mapped onto a lower-dimensional parameter space, convergence toward appropriate solutions is achieved much faster. In addition, as DRC-compliance is intrinsically enforced by the structure of the manifold $G(\mathcal{Z})$, conventional regularization strategies such as minimum feature size rules, curvature constraints, or material boundary penalties are no longer required for fabrication compatibility. Therefore, the optimization search reduces to a direct comparison between the device response and the target response expressed as
\begin{equation}
\arg \min_{z \in \mathcal{Z}} \Big\{ \| F(G(z)) - y^{*} \|^2 \Big\}.
\label{eq:manifold_inv_problem}
\end{equation}
As depicted in Fig.~\ref{fig:figure1}, the intersection $F^{-1}(y^{*}) \cap \mathcal{S}$ corresponds precisely to fabrication-compatible solutions that also satisfy the target response. The feasibility of this optimization in such a lower-dimensional latent space relies on the existence of a well-behaved manifold that ensures fabrication compatibility of any $G(z)$ for a given fabrication platform. In the following section, we discuss how this manifold is constructed through a differentiable convolutional deep neural network architecture.

\section{Intrinsically DRC-Compliant Generative Manifold}

\begin{figure*}[!ht]
    \centering
    \includegraphics[width=\textwidth]{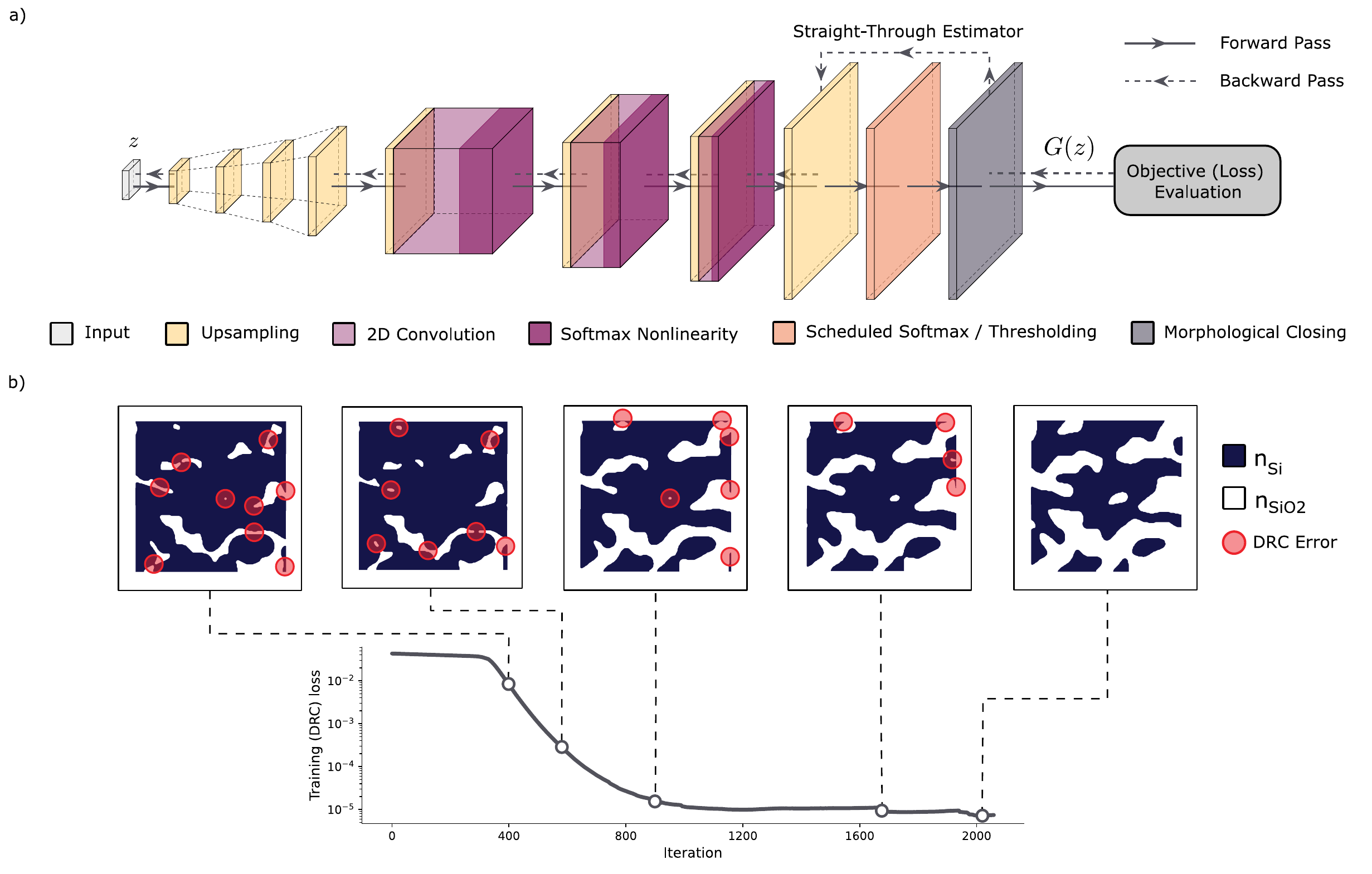}
    \caption{DRC-compliant generative model for electron-beam lithography. 
    (a)~Schematic of the proposed architecture, illustrating the sequence of operations from a latent variable $z$ to the final device layout. The generator comprises four cascaded upsampling stages followed by three upsampling--convolutional layers, with nonlinear transformations including softmax, thresholding, and morphological closing to ensure design-rule compliance. The straight-through estimator (STE) is employed in the backward pass to preserve gradient flow despite the non-differentiable thresholding step. The generated structure $G(z)$ is evaluated through a design objective (DRC loss) and is iteratively optimized. 
    (b)~Training dynamics of the model. The DRC loss decreases from $3 \times 10^{-2}$ to $7 \times 10^{-6}$, representing a reduction by over four orders of magnitude. Insets show representative permittivity distributions at different training stages, with silicon ($n_{\mathrm{Si}}$), silica ($n_{\mathrm{SiO_2}}$), and length-scale violations (red circles) highlighted. As training progresses, violations are systematically eliminated, and the final distribution satisfies DRC constraints with no residual errors.}
    \label{fig:figure2}
\end{figure*}
The goal of the generative manifold is to learn how to map a latent input $z$ to a fabrication-compliant permittivity distribution $G(z)$ that is free of DRC violations, and with smooth/well-defined material boundaries. This is achieved by the sequence of operations depicted by the model in Fig.~\ref{fig:figure2}(a). This model takes as input a low-dimensional, grayscale array with values in the $[0,1]$ range, and processes it through four cascaded upsampling operations, each one enlarging its input by a factor of 1.4. This cascaded upsampling allows for multiple smaller steps to effectively construct a larger composite interpolation filter, allowing for a larger part of the final device to be influenced by a broader set of original input values, and mitigating excessive smoothing typically introduced by a single, large interpolation kernel. The model then applies three successive upsampling-convolution operations with 32, 16, and 1 kernels of size $5 \times 5$, respectively. This approach is more appropriate for creating smooth geometries, as it prevents the emergence of checkerboard artifacts resulting from more typical transposed convolution operations. Each convolution is followed by a softmax activation. Following these three stages are an additional sequence of upsampling, modified softmax, and morphological closing operations. This modified softmax operation is configured with a scheduled slope parameter allowing for it to be updated during training, and is used for gradually guiding the formation of well-defined material boundaries during training. The closing operation performs morphological smoothing that fills small gaps and connects narrow openings over a $3 \times 3$ window to generate the final device geometry. This specific construction of the generator allows for a size-agnostic input-output mapping as all upsampling, convolution, softmax nonlinearity, and morphological closing blocks operate independent of the input dimensions. This means that the model can generate layouts of varying dimensions simply by adjusting the input size. As such, the optimization within the latent space can be performed for any desired target device footprint, simply by selecting the corresponding input dimensions. Because the generator is fully convolutional \cite{Long2015}, the learned 
kernels operate as local, scale-invariant transformations that can be applied to 
inputs of arbitrary spatial dimensions. A single trained generator therefore 
serves all device footprints without retraining; and training complexity is 
independent of the target device size.

For efficient and robust training of this generator model, we use a gradient-based approach using forward and backward passes depicted by solid and dashed arrows through each one of the constituent mathematical blocks in Fig.~\ref{fig:figure2}(a). Operations including convolution, softmax, and most morphological transformations are routinely implemented in a differentiable manner, providing direct access to the derivative information through the backward pass. However, as the final softmax layer with the scheduled slope acts as a non-differentiable thresholding operation by the end of training, this layer requires special consideration for its derivative. To mitigate this, we use a straight-through estimator (STE)~\cite{BengioYoshua2013} to handle this non-differentiable block for the computation of a slightly inaccurate, yet still sufficient estimate of the gradient.

Using this gradient information, we train the generator with a topological loss function that incorporates geometric constraints such as minimum width and spacing. Furthermore, since sharp corners also constitute a design rule violation, the loss implicitly includes a curvature constraint to ensure all generated edges are smooth and fabrication-compatible relative to the specified minimum feature size. The upsampling layers in the generator further reinforce this curvature constraint by expanding small feature maps into larger ones while also performing spatial smoothing. Using a single-etch silicon photonic platform with boundaries between $\mathrm{Si}$ and $\text{SiO}_{2}$ materials, we formulate this topological loss~\cite{HammondAlec2021} as
\begin{equation}
\begin{split}
\mathcal{L} = \sum_{i,j} &I_{i,j}^{LW}(\rho) \cdot [\min\{(\rho_{i,j} - \eta_{e}), 0\}]^2 \\
&+ I_{i,j}^{LS}(\rho) \cdot [\min\{(\eta_{d} - \rho_{i,j}), 0\}]^2.
\end{split}
\label{eq:custom_loss}
\end{equation}
Here, $\rho$ represents a normalized permittivity distribution between 0 ($\text{SiO}_{2}$) and 1 ($\text{Si}$), while the scalars $\eta_{d}$ and $\eta_{e}$ serve as permittivity thresholds (between 0 and 1) for the erosion and dilation operations. The auxiliary functions $I_{\text{LW}}$ and $I_{\text{LS}}$ represent inflection regions that determine whether the design topology remains invariant under a strategically chosen sequence of erosions and dilations, see Supporting Information~1. As constructed, this loss function does not explicitly include terms for minimum area or minimum enclosed area. Instead, the generator automatically removes small islands through the differentiable morphological closing layer, whose kernel size exceeds the minimum feature size. Likewise, smoothing introduced by the initial cascaded upsampling prevents the formation of small holes, thereby ensuring compliance with the minimum enclosed area requirement. Importantly, the minimum area and minimum enclosed area thresholds are not rigidly coupled to these minimum linewidth and linespacing parameters, and can be controlled independently by adjusting the upsampling ratio and the morphological closing kernel size. However, the closing kernel's physical footprint must remain above the configured minimum linewidth to avoid removal of otherwise compliant narrow geometrical features. A conservative lower bound on the minimum area and minimum enclosed area can be estimated
directly from the morphological closing kernel operating on the 25 nm output grid: the
$3\times3$ kernel of the EBL generator yields a footprint floor of $75~\text{nm} \times
75~\text{nm}$, and the $7\times7$ kernel of the PL generator yields $175~\text{nm} \times
175~\text{nm}$.

To ensure that design rule constraints are satisfied not only within the design  domain but also at the device boundaries and at waveguide--device junctions, two 
complementary measures are incorporated during training and inference. During training, a $1~\mu$m (40-pixel) SiO$_2$ padding is appended to the top, bottom, and right edges of every generated device, and input/output waveguides with randomly assigned widths and positions along these edges are incorporated prior 
to morphological closing, explicitly exposing boundary regions and waveguide 
interfaces to the DRC loss. During inference, the input and output waveguides 
are similarly integrated into the binary permittivity distribution immediately 
after thresholding and before morphological closing, so that any DRC violations 
arising at the waveguide--device interface are resolved by the closing operation 
before the layout is converted to a GDS file. The complete boundary compliance 
procedure is described in detail in Supporting Information~2. For this specific generator and loss function, we specify minimum width and spacing as 60~nm, for compatibility with standard EBL processes available. The model also allows for configuration with larger feature sizes for PL, as discussed in the later sections.

To train this generator, we created a custom dataset of inputs by representing 2D noisy latent space distributions using Perlin noise, as it produces smooth yet stochastic variations across multiple length scales~\cite{Caseman2015}. Compared to pixel-based uncorrelated random noise, this approach yields a highly diversified training distribution by capturing both pixelated, spatially high frequency patterns as well as inputs with large geometrical features. To incorporate diversity throughout the dataset, we generated 1000 unique samples by randomly varying the scale, offset, and noise dimensionality, thereby obtaining multiple granularities across different topologies for improving the generalization capabilities of the generator. These parameters control the relative sizes of textures, spatial shifts and alignment of features, and number of independent stochastic components that define each pattern. For more details on each one of these parameters, and visual comparison of the latent space samples, see Supporting Information~3. We constrained each input distribution to lateral dimensions between 4 and 45 pixels in each direction, such that the dataset covers multiple physical device sizes between 1~$\mu$m and 20~$\mu$m, after the generator upscaling. The input/output mapping between pixels in the latent space and physical dimensions in the device space are provided in Supporting Information~4.

The generator was trained using an ADAM optimizer as shown in Fig.~\ref{fig:figure2}(b). The goal of this training procedure is to effectively shape $G(\mathcal{Z})$ according to the loss function specified above, such that the output devices are all intrinsically fabrication-compliant. The generator model begins with a DRC loss of $4 \times 10^{-2}$, decreases slowly over the first 300 iterations, and falls further as the generator outputs sharpen, ultimately reaching $8 \times 10^{-6}$ after 2060 iterations (see Methods for more training details). With this progression, the number of DRC violations for generated devices also gradually decreases. This is illustrated by snapshots of the output $G(z_{0})$ provided throughout the training, for a specific input $z_{0}$. The training procedure gradually modifies the generator parameters to smooth and correct regions violating minimum length and spacing constraints, eliminating length-scale violations in the process. At the end of training, the manifold $G(\mathcal{Z})$ is effectively a subset of $\mathcal{S}$ (the set of all fabrication-compliant devices), ultimately producing an intrinsically DRC-compliant model where all inputs in the latent space correspond to physically DRC-clean device layouts.

During the inference of the trained generator, the scheduled softmax is replaced with a thresholding operation to obtain a discrete layout. Since this thresholding layer is non-differentiable, its gradient is approximated using the STE introduced earlier, allowing the entire generator and pipeline to remain differentiable and enabling gradient-based optimization of the latent space.

\begin{figure*}[!ht]
    \centering
    \includegraphics[width=\textwidth]{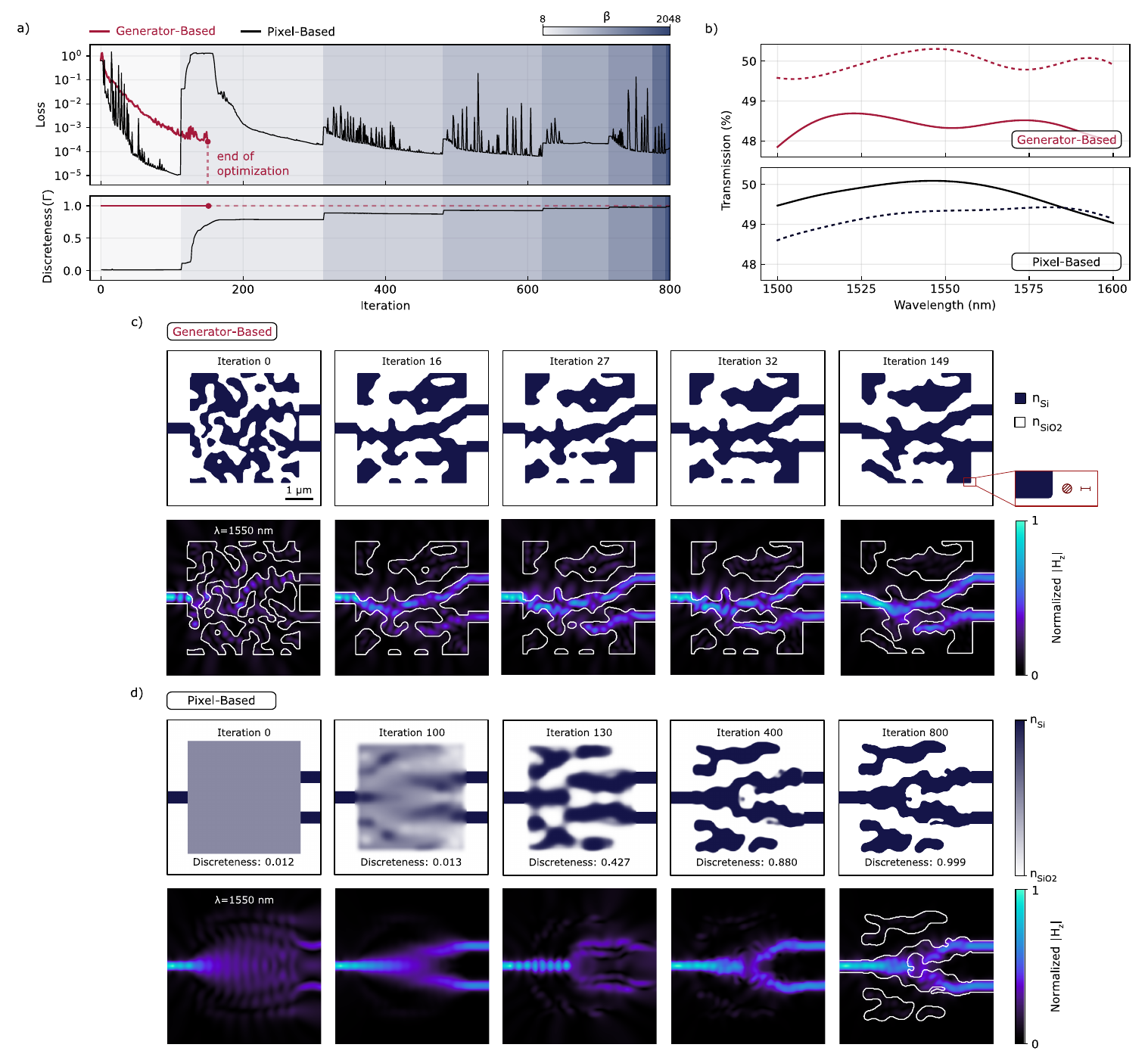}
    \caption{Comparison of generator-based and pixel-based inverse design for 50/50 power splitters designed for the electron beam lithography platform. (a)~Evolution of the loss function and $\Gamma$ during optimization. The generator-based method (red) converges to a loss of $2.8 \times 10^{-4}$ within 149 iterations while maintaining full discreteness of 1 throughout. In contrast, the pixel-based method (black) reaches a lower loss of $1.3 \times 10^{-4}$ after 800 iterations, but requires progressive increases in the projection strength parameter $\beta$ (color-coded, from 8 to 2048) to approach discretization, corresponding to over a 5-fold reduction in computational cost with the generator-based formulation. (b)~Transmission spectra of the optimized devices, where solid and dashed lines denote the bottom and top output ports, respectively. (c)~Generator-based optimization: snapshots of the permittivity distribution (top) and the corresponding normalized magnetic field $\lvert H_z \rvert$ at $\lambda = 1550$~nm (bottom) across representative iterations, inset shows minimum area (circle) and length (bar) constraints The white contours denote the silicon boundaries. The design remains strictly discrete at every iteration. (d)~Pixel-based optimization: analogous snapshots of permittivity (top) and $\lvert H_z \rvert$ fields (bottom), illustrating gradual discretization that is only fully achieved at the final iteration. $\Gamma$ in each snapshot quantitatively tracks this transition.}
    \label{fig:figure3}
\end{figure*}

\clearpage

\section{Device Optimizations}
In this section, we demonstrate the versatility of our generator by designing several representative classes of photonic devices including power splitters, spectral mux/demux, and a mode converter. The device optimization pipeline is constructed through modular computational blocks. For all devices, the generator first creates the device's geometry $G(z)$ from the latent space representation $z$. This is followed by a differentiable electromagnetic simulator and the final objective (loss) evaluation based on the specified target functionality. An optimizer then iteratively updates the latent space representation $z$ according to this objective evaluation. This modular construction allows users to utilize any differentiable electromagnetic simulator that suits their application~\cite{HughesTyler2019, SchubertFrederik2025, HootenSean2025, TangRuiJie2023}. For our specific applications, we employ a factorization-cached implementation of the Finite-Difference Frequency-Domain (FDFD) solver, which stores and reuses the matrix factorizations of the discretized Maxwell's equations throughout the iterative design process~\cite{Dasdemir2023}. The optimizations are carried out over a broadband wavelength range from 1500~nm to 1600~nm using a single-etch, 220-nm-tall Silicon-on-Insulator (SOI) platform. For the power splitters and spectral mux/demux devices, we use the fundamental transverse electric modes ($\text{TE}_{0}$) of 450-nm-wide input and output waveguides. The mode converter features a larger output waveguide width (750~nm) to support a guided $\text{TE}_{1}$ mode. This entire pipeline is designed to confine the device optimization procedure to a low-dimensional, fabrication-aware subspace. As a result, the generated final devices satisfy the objective photonic functionality as closely as possible, and are free of any DRC violations, including at the device boundaries and at the waveguide--device junction regions (see Supporting Information~2 for the complete boundary compliance and waveguide integration procedure).

\subsection*{Design of Broadband Power Splitters with Variable Splitting Ratios}

To place our proposed approach within the existing inverse design domain, we first compare it against conventional, pixel-based inverse design methods by designing a compact, broadband 50/50 power splitter as a benchmark device. The dimensions of this benchmark device are identical at $4.2 \times 4.2~\mu\text{m}^{2}$ for both our proposed method and the conventional approach. The target condition is that the optical power is evenly distributed between the two output waveguides within the 1500--1600~nm wavelength range. In conventional inverse design, the device geometry is represented by individual grayscale pixels, which are iteratively updated during optimization. Operations such as averaging and projection operations are applied over the pixel field to drive pixel values towards 0 or 1 for enforcing discretization and creating distinct material boundaries. These operations are typically updated through pre-determined or adaptive schedules, effectively redefining the optimization problem with each iteration~\cite{LazarovBoyan2016, WangFengwen2022, Molesky2018, RasmusChristiansen2021, Jensen2011, HammondAlec2021}. The scheduled update of the discretization projection strength often introduces abrupt changes in the objective function, and requires further optimization for the device performance to be recovered~\cite{HammondAlec2021, VercruysseDries2019, ChenYuchen2023}. This behavior is illustrated by the black curves in Fig.~\ref{fig:figure3}(a), where successive increases in the objective correspond to scheduled increments of the discretization strength parameter ($\beta$). This parameter controls the slope of a sigmoid function used to drive individual pixels towards $\text{SiO}_{2}$ (0) or $\text{Si}$ (1) (see Methods). Introducing strong discretization too early in the optimization can lead to difficult convergence, entrapment in local minima with insufficient performance, and a significantly slower overall process~\cite{MichaelsAndrew2018, GuestJK2004, ChenYuchen2023}. Although gradual discretization circumvents these issues, it also requires a substantial number of iterations to explore grayscale, unrealizable geometries with ill-defined material boundaries, thereby limiting both optimization efficiency and practical utility. For this device, convergence is achieved with a final objective of $1.3 \times 10^{-4}$ in 800 iterations (see Methods for convergence conditions), after which the objective does not significantly improve. The discreteness metric $\Gamma$, (defined as the portion of black \& white pixels, see Supporting Information~5) begins at 0.01, corresponding to a nearly fully grayscale device, and increases proportionally with projection-strength updates until it approaches 1, representing well-defined material boundaries.

In contrast, the generator-based approach inherently constrains the search to fabrication-compliant geometries, allowing for a much more efficient exploration of the design space while inherently satisfying design rule constraints. Unlike pixel-based approaches, this method requires no scheduled projection-strength updates, and therefore avoids the stepwise increases in the objective function that such scheduled updates periodically introduce. As shown by the red curves in Fig.~\ref{fig:figure3}(a), it maintains well-defined material boundaries throughout the entire optimization. For a device with identical dimensions, the optimization is performed in a latent space of dimension 18 × 18, and convergence is reached in 149 iterations, over 5-times faster than the conventional method, with a final objective of $2.8 \times 10^{-4}$. Unlike the gradually discretized device in traditional inverse design, the gradient-based approach maintains a fully discretized geometry throughout the entire optimization process, as indicated by the constant $\Gamma$ at 1.

In our generator-based approach, the underlying optimization framework remains structurally identical to conventional inverse design, with the exception of inference through the trained generator. These added steps only marginally increase the per-iteration computational requirements for two key reasons. Firstly, modern Graphics Processing Units (GPUs) are highly optimized for parallel processing of image-based operations, including the convolutions and point-based nonlinearities used in the generator architecture. This inherent parallelism ensures the generator inference remains computationally efficient. More importantly, in both conventional and generator-based methods, the physical device simulations consume the vast majority of the computational resources for each iteration~\cite{PeterWiecha2021, ChenHao2024, HootenSean2025}. As a result, our generator-based design process offers a highly favorable trade-off resulting in a significantly reduced iteration count with marginal additional computational overhead. Crucially, the generator-based optimization process requires no regularization or objective scheduling to enforce material boundaries. This elimination of scheduled parameter updates also streamlines the design process, making it significantly more robust and usable for designers, with no need for repeated parameter tuning, optimization restarts, or manual intervention. These differences highlight the generator-based methodology as a more efficient, autonomous, and parameter-independent inverse design platform compared to conventional methods.

Fig.~\ref{fig:figure3}(b) illustrates the transmission spectra for both optimized devices, with dashed and solid lines representing the top and bottom output ports, respectively. Both approaches yield devices with good performance, maintaining a near-ideal 50/50 power-splitting ratio across the entire 1500--1600~nm wavelength range. Fig.~\ref{fig:figure3}(c) and Fig.~\ref{fig:figure3}(d) present the final device geometries and the normalized $H_{z}$ field profiles at 1550~nm, as obtained through Finite-Difference Frequency-Domain (FDFD) simulations performed on the two devices. Inset includes length scale and area constraint. The circle indicator corresponds to the minimum area and enclosed area constraint (circle with a diameter of 75 nm for EBL and 150 nm for PL), while the line bar next to it corresponds to 60 nm for EBL and 150 nm for PL, which corresponds to the minimum linewidth and minimum linespace constraint. DRC procedures confirm that both final geometries fully satisfy the fabrication constraints in addition to their broadband, efficient power splitting functionality. Crucially, the generator-based device maintains its well-defined material boundaries throughout all optimization iterations. This approach eliminates the requirement for gradual increases in discretization strength, which is typically necessary for the convergence of conventional grayscale designs.

\begin{figure}[!h]
    \centering
    \includegraphics[width=\columnwidth]{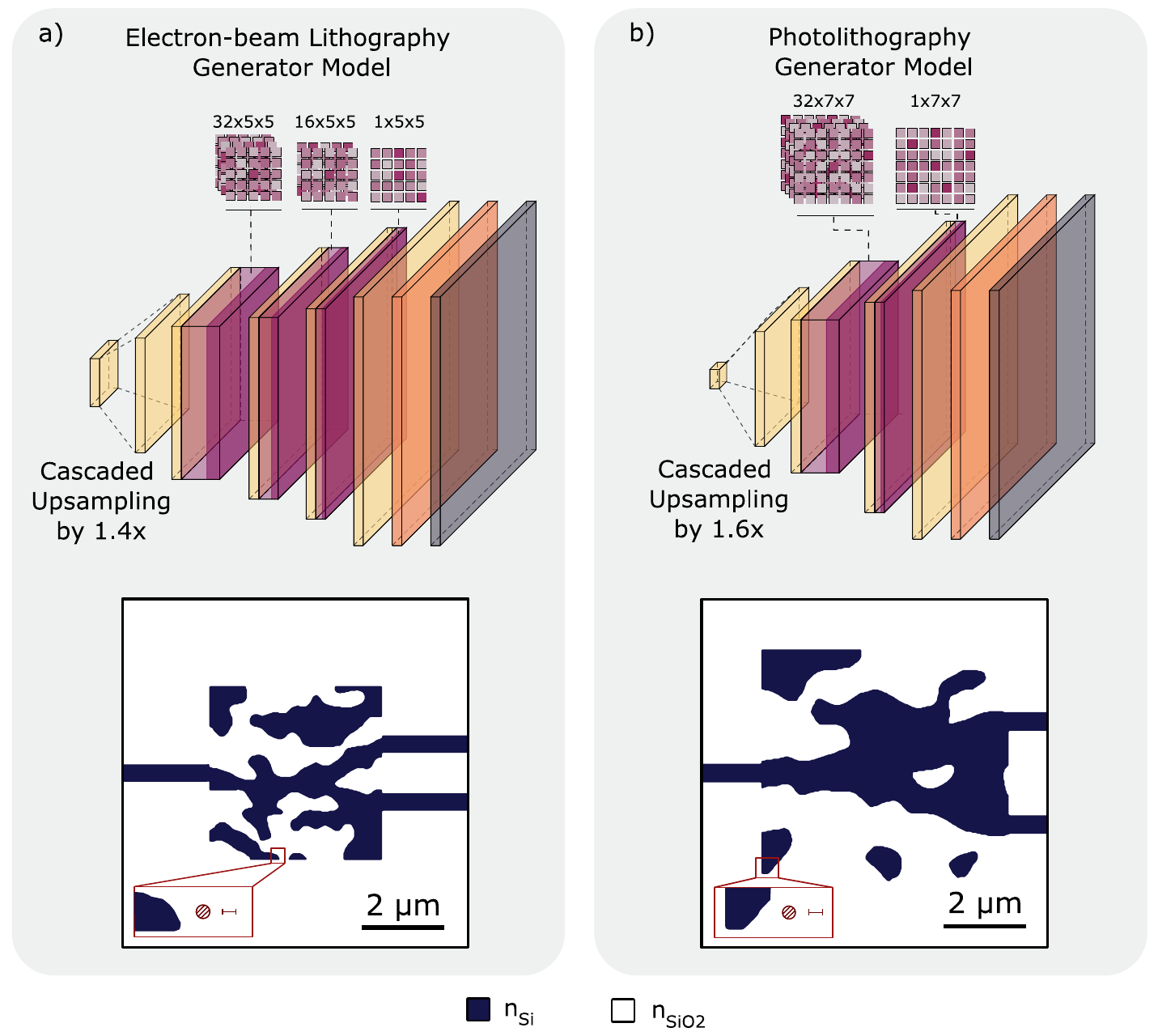}
    \caption{(a) Architectures of generator models designed for two different lithography models. The electron-beam lithography (EBL) model employs a cascaded upsampling architecture with four 1.4$\times$ upsampling stages. This architecture incorporates three upsampling-convolution layers with varying filter sizes (32, 16, and 1) and a $5\times5$ kernel. The output is refined using thresholding and morphological closing layers. (b) The photolithography (PL) model features a similar cascaded upsampling architecture, but with four 1.6$\times$ upsampling stages and two upsampling-convolution layers with a 16.1 filter size and a $7\times7$ kernel. The bottom row of the figure displays the permittivity distributions of a 50/50 power splitter device generated by each model, with footprints of $4.2 \times 4.2~\mu\text{m}^2$ (EBL) and $6 \times 6~\mu\text{m}^2$ (PL). The models are constrained by minimum feature sizes of 60~nm for EBL and 150~nm for PL. The dark blue and white regions represent silicon ($n_{\mathrm{Si}}$) and silicon dioxide ($n_{\mathrm{SiO_2}}$), respectively. Insets show minimum area (circle) and length (bar) constraints.}
    \label{fig:figure4}
\end{figure}

\begin{figure*}[ht!]
    \centering
    \includegraphics[width=\textwidth]{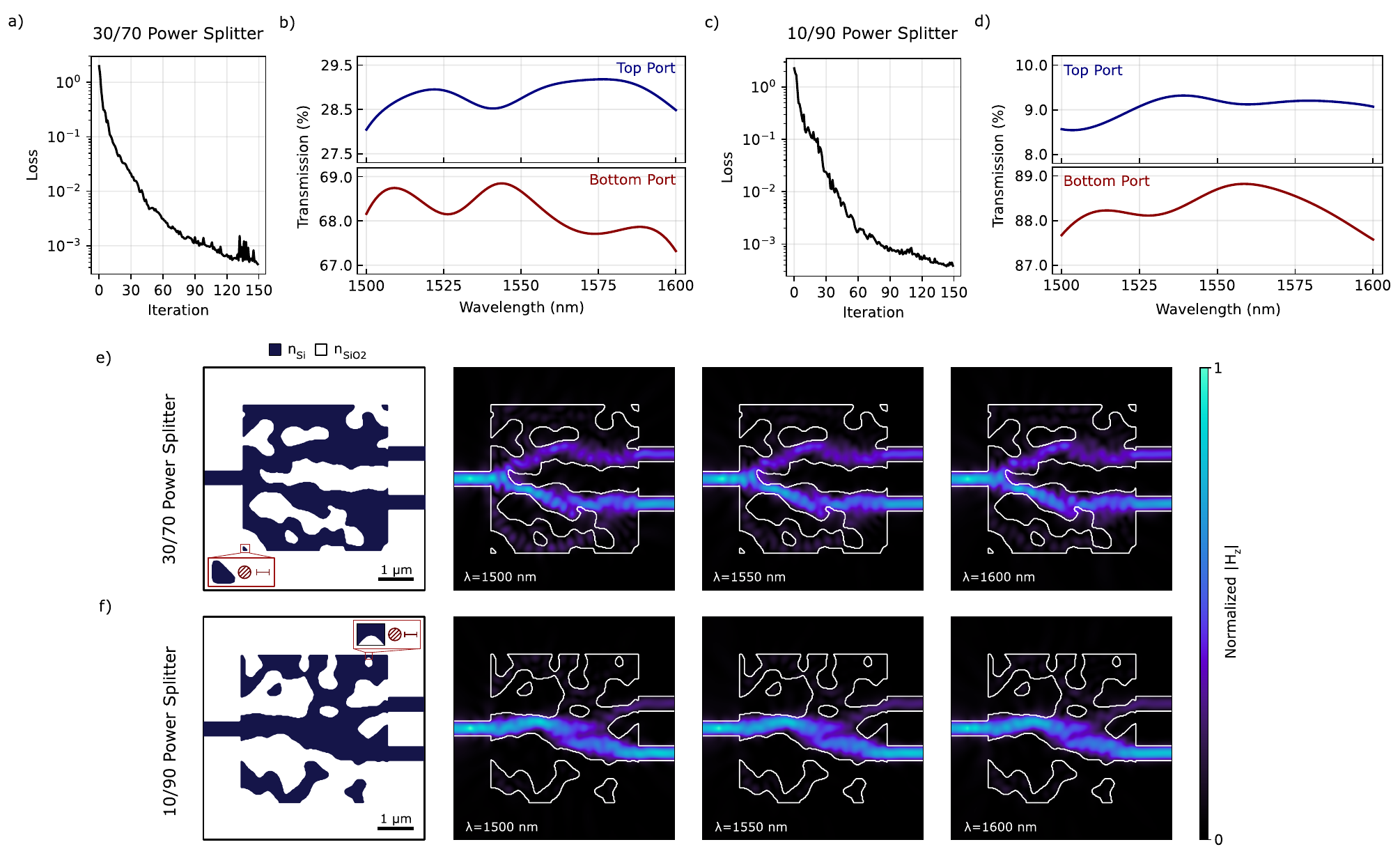}
    \caption{Ultrabroadband 70/30 and 90/10 power splitter results designed with the electron-beam lithography generator model. 
    (a)~Optimization loss curve for the 30/70 power splitter, showing the loss decreasing to $5 \times 10^{-4}$ over 150 iterations. 
    (b)~Transmission curves for the top and bottom ports of the 30/70 power splitter, demonstrating stable performance across the $1500$--$1600$~nm wavelength range. 
    (c)~Optimization loss curve for the 10/90 power splitter, converging to a loss of $4 \times 10^{-4}$ after 150 iterations. 
    (d)~Transmission curves for the top and bottom ports of the 10/90 power splitter. 
    (e)~Permittivity distribution of the 30/70 power splitter (left, with silicon in dark blue and SiO$_2$ in white), alongside the normalized magnetic field ($H_z$) distribution at wavelengths of 1500, 1550, and 1600~nm. The wavelength-insensitive field profiles indicate that the splitting ratio remains constant across the simulated spectrum. 
    (f)~Permittivity distribution and normalized magnetic field distribution for the 10/90 power splitter, corresponding to the same wavelengths as in (e). Insets show minimum area (circle) and length (bar) constraints.}
    \label{fig:figure5}
\end{figure*}

Our generator-based approach provides a general platform for optimization for compatibility with multiple different fabrication methods and foundry standards. We illustrate this by training two separate generators, corresponding to two widely used fabrication technologies: EBL (with $60\,\text{nm}$ min. feature size) and PL (with $150\,\text{nm}$ min. feature size) as shown in Fig.~\ref{fig:figure4}. The larger feature size necessary for the PL generator model was achieved through a greater scaling ratio of 1.6 for each initial upsampling stage, one fewer upsampling-convolution block compared to the e-beam model, and larger convolution kernels of $7 \times 7$ in size. Achieving DRC compliance at larger feature sizes requires larger convolutional kernels, effectively expanding the receptive fields of the convolutional layers. To manage the increased precision required for minimum-width and enclosed-area rules at this lower feature size, the $60\text{-nm}$ model utilizes three upsampling convolution layers compared to only two in the $150\text{-nm}$ version. The training of the PL generator model is detailed in Supporting Information~6. The output grid for both the e-beam and photolithography generators were set to a pixel size of $25\,\text{nm} \times 25\,\text{nm}$. The electromagnetic simulation grid used by the FDFD solver was also set to 
$25~\text{nm} \times 25~\text{nm}$, matching the generator output resolution. 
This resolution was used consistently across all device types, including power 
splitters, wavelength duplexers, and mode converters. However, the generator 
output can also be resampled to other simulation grids via interpolation without 
retraining from scratch. A statistical validation of DRC compliance across a wide range of device footprints for both generator models is provided in Supporting Information~7.

The two 50/50 splitters designed respectively with the two models intrinsically compliant with e-beam and photolithography are shown in the bottom half of Fig.~\ref{fig:figure4}. The first device shown here is the same as the one in Fig.~\ref{fig:figure3}, and is displayed here for comparison. The second device is the output of the photolithography-compliant generator described above. When using this PL model, we expand the device footprint from $4.2 \times 4.2~\mu\text{m}^{2}$ to $6 \times 6~\mu\text{m}^{2}$. This larger design region is necessary because the PL model, constrained by a larger minimum feature size compared to the e-beam model, is less capable of supporting the same design performance within the smaller footprint. This optimization for this device converges after 150 iterations, yielding a final loss value of $2 \times 10^{-3}$. The higher loss compared to the e-beam generator model arises from the stricter, larger fabrication constraint of 150~nm. This model is more constrained, and is therefore unable to introduce finer structural details. This reduces the design flexibility, and makes the design task a more challenging problem. Consequently, the lowest attainable local minimum stabilizes around a slightly higher loss, compared to $3.2 \times 10^{-4}$ obtained with the e-beam-compliant generator. The number of iterations required for convergence remains similar between the two cases since both optimizations operate within comparable latent space dimensions. The transmission spectra, details of the device geometries, and FDFD field plots of the photolithography-compliant power splitter are provided in Supporting Information~8.

Both generator models are inherently size-agnostic due to their fully 
convolutional architecture. Since all operations in the generator (including 
upsampling convolutions, pointwise nonlinearities, and morphological closing) 
operate locally, they all process a fixed number of neighboring pixels. This 
means that the same trained generator model can process latent inputs of 
arbitrary dimensions without requiring retraining. For all presented results 
here, a single generator model is trained once for each fabrication technology 
(EBL or PL), and reused with different footprints by varying the input latent 
space dimension. This latent space dimension for each device is determined based 
on the desired physical footprint, with larger footprints utilizing 
higher-dimensional latent spaces to provide additional degrees of freedom. This 
flexibility allows the optimization to explore richer design spaces while 
maintaining strict adherence to fabrication constraints throughout the entire 
design trajectory. Beyond these designs, the framework can optionally enforce mirror symmetry by partitioning
the latent space along the length-wise axis and mirroring the permittivity map at each
forward pass. This guarantees in-phase outputs by construction and also reduces the
effective latent dimensionality by half. This capability integrates into the existing
optimization pipelines without any retraining or additional overhead. The symmetric power
splitter designed using the EBL-compatible manifold is demonstrated in Supporting Information 9, with a simulated insertion loss of 0.067 dB at 1550 nm. This device reaches convergence in 110 iterations, fewer than the 149 iterations required for
the non-symmetric counterpart, while maintaining matching output transmissions across the
full 1500--1600~nm wavelength range.

To demonstrate the utility and generalization capabilities of our generator models, we design power splitters with arbitrary splitting ratios. Specifically, we demonstrate two additional power splitters with splitting ratios of 30/70 and 10/90. Both devices share the same footprint as the initial 50/50 power splitter, namely $4.2 \times 4.2~\mu\text{m}^{2}$, and are optimized using the same latent space dimension of 18 × 18. For all three power splitters (50/50, 30/70, and 10/90), the output waveguide 
separation is designed to be $1.5~\mu$m, measured center to center between the 
two output waveguides. For the 30/70 power splitter, we select six uniformly-spaced wavelengths between 1500~nm and 1600~nm and set target operation at all wavelengths for a top port transmission of 30\% and a bottom port transmission of 70\%. As shown in Fig.~\ref{fig:figure5}(a), the optimization reduces the loss to $5 \times 10^{-4}$ after 150 iterations. Fig.~\ref{fig:figure5}(b) shows that the transmission varies between 27\% and 29\% at the top port, while it remains between 67\% and 69\% at the bottom port. The simulated insertion loss of this device is 0.11~dB at 1550~nm. For the 10/90 power splitter, we again select the same six wavelengths, targeting 10\% transmission at the top port and 90\% at the bottom port. After 150 iterations, the optimization converges to a minimum loss of $4 \times 10^{-4}$, as illustrated in Fig.~\ref{fig:figure5}(c). Fig.~\ref{fig:figure5}(d) shows that the top port transmission between 8\% and 10\%, while the bottom port transmission varies between 87\% and 89\%. The simulated insertion loss of this splitter is approximately 0.09~dB. In both optimizations, the local minimum emerges after 150 iterations, similar to the 50/50 power splitter. This consistency can be attributed to the identical latent space dimensions across the three cases. Fig.~\ref{fig:figure5}(e) presents the final geometry of the optimized 30/70 splitter, along with the normalized magnetic field magnitude from finite-difference frequency-domain (FDFD) simulations at three different wavelengths, illustrating how the input power distributes across the output ports. Similarly, Fig.~\ref{fig:figure5}(f) shows the geometry of the optimized 10/90 splitter and its normalized magnetic field magnitudes at three wavelengths, confirming the designed splitting behavior. These simulated field distributions verify the transmission behavior and the broad operation bandwidth of the optimized devices.

\subsection*{Multi-Wavelength Photonic Duplexer}

\begin{figure}[!h]
    \centering
    \includegraphics[width=\columnwidth]{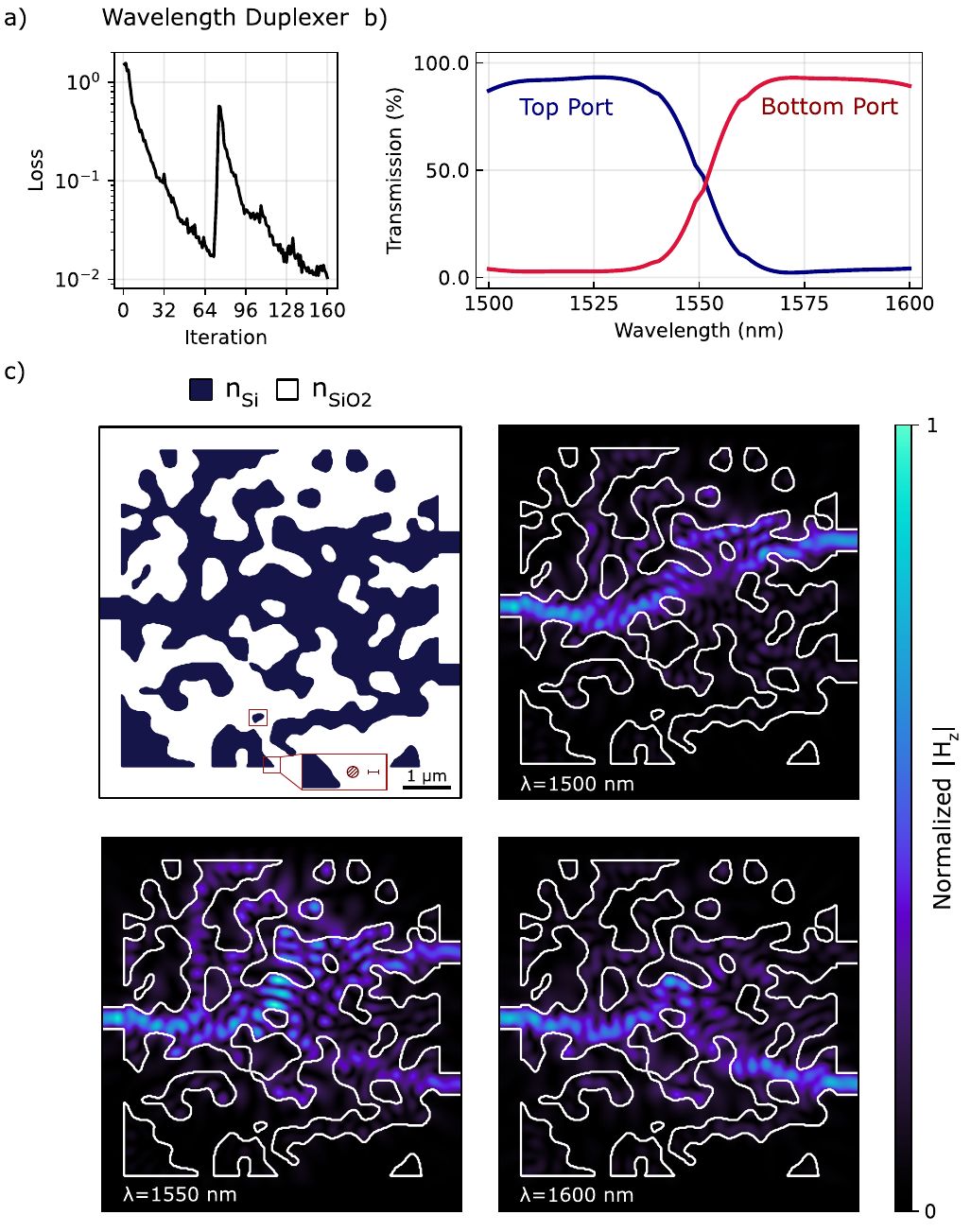}
    \caption{Photonic duplexer designed with the electron-beam-lithography-compliant generator model. 
    (a)~Optimization loss curve of the wavelength demultiplexer, showing a rapid decrease to $1 \times 10^{-2}$ after 160 iterations. 
    (b)~Transmission spectra of the optimized duplexer, demonstrating efficient separation between the top and bottom output ports with peak transmission of 94\% (corresponding to an insertion loss of 0.25~dB). 
    (c)~Final device layout (top left), where the permittivity distribution is represented with silicon in dark blue and SiO$_2$ in white, inset shows minimum area (circle) and length (bar) constraints. The device satisfies all design rule constraints. The accompanying field maps illustrate the normalized FDFD magnetic field distributions at 1500, 1550, and 1600~nm, confirming robust duplexing functionality across the operating wavelength range.}
    \label{fig:figure6}
\end{figure}

To further demonstrate the capability of our proposed model, we design a broadband silicon photonic wavelength duplexer as an additional benchmark device. This device routes input light within the $1500$--$1550\,\text{nm}$ wavelength range to the top port, while signals in the $1550$--$1600\,\text{nm}$ wavelength range are directed to the bottom port, effectively realizing short-pass and long-pass output ports. The footprint of the device is $7 \times 7\,\mu\text{m}^2$, with a corresponding latent space dimension of $28 \times 28$. For the optimization objective, we use ten target wavelengths between $1500\,\text{nm}$ and $1600\,\text{nm}$.  The first five wavelengths are set to achieve 100\% transmission into the top port, while the remaining five wavelengths are targeted for 100\% transmission into the bottom port.

After 160 optimization iterations, a minimum loss of $1 \times 10^{-2}$ is achieved, as shown in Fig.~\ref{fig:figure6}(a). The abrupt jump in the loss function observed around iteration 70 is a direct consequence of the optimizer’s trajectory as it navigates the complex loss landscape of the latent space. This discontinuity suggests that the optimizer becomes temporarily trapped in a local minimum, and subsequently transitions to a more favorable region of the search space. As illustrated in Fig.~\ref{fig:figure6}(b), the device successfully performs the spectral duplexing operation between the top and bottom ports. The achieved simulated transmission reaches approximately 94\%, which corresponds to an insertion loss of only $0.25\,\text{dB}$. The output waveguide separation for this wavelength duplexer is $3~\mu$m, 
measured center to center between the two output waveguides. The geometry of the obtained wavelength demultiplexer reveals the distribution of $\text{SiO}_2$ and Si materials, with a successful DRC verification. Fig.~\ref{fig:figure6}(c) plots the normalized FDFD magnetic field distributions at $1500$, $1550$, and $1600\,\text{nm}$, which clearly demonstrate how the device routes input signals to the corresponding output ports across different wavelengths.

\subsection*{Broadband Photonic Mode Converter}

\begin{figure}[!h]
    \centering
    \includegraphics[width=\columnwidth]{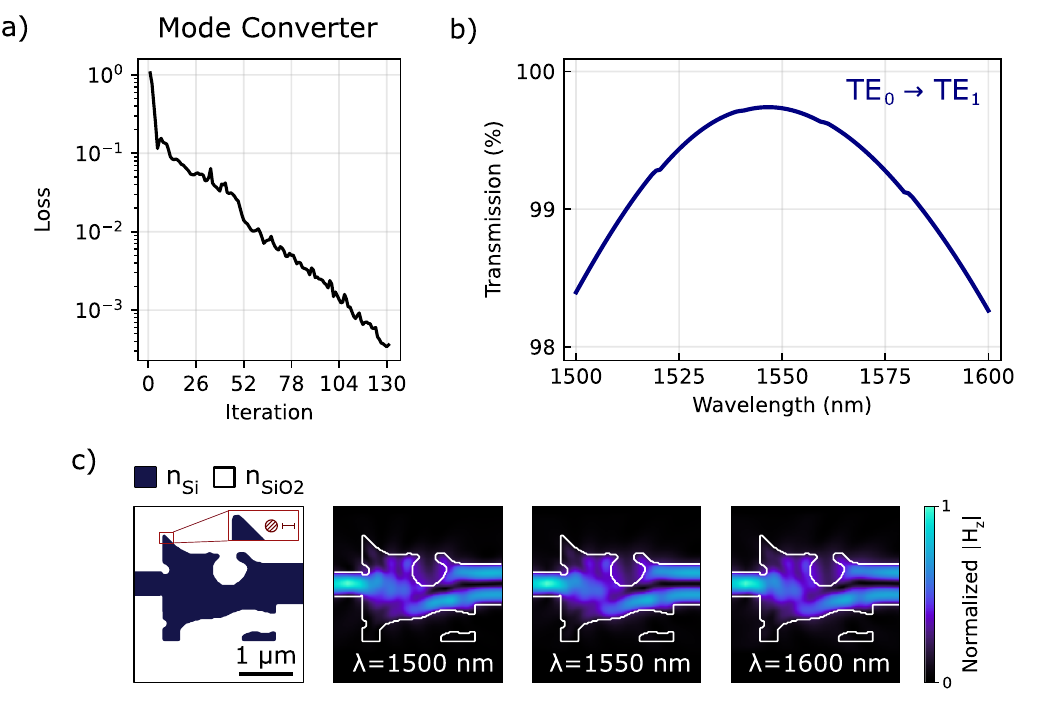}
    \caption{Results of the broadband photonic mode converter designed with the electron-beam-lithography-compliant generator model. (a)~Convergence of the optimization loss to $3 \times 10^{-4}$ over 130 iterations. 
    (b)~The simulated transmission curve for the mode converter, showing a $\mathrm{TE}_0$ to $\mathrm{TE}_1$ conversion ratio between 98\% and 100\% across the wavelength range of $1500$--$1600$~nm. The simulated conversion efficiency reaches $\sim 99$\% at 1550~nm, corresponding to an insertion loss of only 0.03~dB. 
    (c)~The optimized device permittivity distribution (left) and the simulated normalized magnetic field distributions ($\lvert H_z \rvert$) at input wavelengths of 1500, 1550, and 1600~nm. Inset shows minimum area (circle) and length (bar) constraints.}
    \label{fig:figure7}
\end{figure}

Using the electron-beam-lithography-compliant generator model, we also design a 
broadband mode converter whose purpose is to transform the fundamental $\text{TE}_0$ 
mode at the input into the $\text{TE}_1$ mode at the output. Since this task is relatively simple compared to other devices, we use a significantly smaller design footprint at $2.1 \times 2.1\,\mu\text{m}^2$. In the optimization, we set the objective to achieve a 100\% overlap with the $\text{TE}_1$ electric field distribution at the output port. As shown in Fig.~\ref{fig:figure7}(a), the loss reaches a minimum of $3 \times 10^{-4}$. In this case, the optimization is carried out in a latent space of dimension $10 \times 10$, which is sufficient for the device to converge to a local minimum within 130 iterations. Fig.~\ref{fig:figure7}(b) illustrates that transmission to the $\text{TE}_1$ output remains between 98\% and 100\%, with a simulated insertion loss of only $0.03\,\text{dB}$ at $1550\,\text{nm}$. The device geometry is presented in Fig.~\ref{fig:figure7}(c) with additional field plots at wavelengths of $1500$, $1550$, and $1600\,\text{nm}$ verifying broadband and near-lossless operation in a compact footprint. To verify robustness to fabrication imperfections, all devices were simulated under $\pm 15$~nm over-etching and under-etching conditions. Transmission 
spectra across all device classes and both platforms remain within acceptable 
bounds of the as-designed performance (see Supporting Information~10).

\begin{table*}[!ht]
\centering
\begin{ruledtabular}
\begin{tabular}{lccccccccccc}
Reference 
& \multicolumn{3}{c}{Power Splitter} 
& \multicolumn{4}{c}{Wavelength Duplexer} 
& \multicolumn{3}{c}{Mode Converter} 
& \multicolumn{1}{c}{Minimum} \\
\cline{2-4} \cline{5-8} \cline{9-11}
& IL & Footpr. & Iter. 
& IL & Footpr. & ER & Iter. 
& IL & Footpr. & Iter. 
& Feature \\

& (dB) & ($\mu$m$^2$) & 
& (dB) & ($\mu$m$^2$) & (dB) &
& (dB) & ($\mu$m$^2$) & 
& (nm) \\
\hline
Wang \emph{et al.}~\cite{KaiyuanWang2020}
& 0.32 & $2.6 \times 2.6$ & 50 
& -- & -- & -- & -- 
& -- & -- & -- 
& 70 \\
Chen H. \emph{et al.}~\cite{ChenHao2024}
& 0.24 & $3.0 \times 2.6$ & 75 
& 0.62 & $3.0 \times 3.0$ & 12 & 150 
& 0.24 & $2.6 \times 3.0$ & 100 
& 70-110 \\
Schubert \emph{et al.}~\cite{SchubertMartin2022} 
& 0.34 & $3.2 \times 2.0$ & 150 
& 1.17 & $6.4 \times 6.4$ & 20 & 150 
& 0.45 & $1.6 \times 1.6$ & 100 
& 100 \\
Chen Y. \emph{et al.}~\cite{ChenYuchen2023}
& 0.46 & $2.8 \times 2.8$ & 450 
& -- & -- & -- & -- 
& -- & -- & -- 
& 130 \\
Hiesener \emph{et al.}~\cite{HiesenerJacob2025}
& -- & -- & -- 
& -- & -- & -- & -- 
& 0.64 & $6.0 \times 3.0$ & 120 
& 85* \\
Our work (EBL)
& \textbf{0.06} & $4.2 \times 4.2$ & 150 
& \textbf{0.25} & $7.0 \times 7.0$ & 16 & 160 
& \textbf{0.03} & $2.1 \times 2.1$ & 130 
& 60 \\
Our work (PL)
& \textbf{0.22} & $4.2 \times 4.2$ & 150 
& \textbf{1.19} & $7.0 \times 7.0$ & 12 & 160 
& \textbf{0.10} & $2.1 \times 2.1$ & 130 
& 150 \\
\end{tabular}
\end{ruledtabular}
\caption{Performance comparison of the proposed DRC-compliant inverse design method with previously reported simulation results from approaches for power splitters, wavelength demultiplexers, and mode converters. Device footprints, electromagnetic solvers, 
and optimization platforms vary across the compared results. (Minimum feature 
constraints: \cite{KaiyuanWang2020}~-- discrete building-block library with ${\sim}70$~nm 
effective minimum etched feature; \cite{ChenHao2024}~-- 70~nm (wavelength multiplexer) 
and 110~nm (power splitter, mode converter); \cite{SchubertMartin2022}~-- 100~nm; 
\cite{ChenYuchen2023}~-- 130~nm minimum radius of curvature; \cite{HiesenerJacob2025}~-- commercial 
foundry with proprietary design rules. Our work~-- 60~nm minimum feature size 
for EBL, 150~nm minimum feature size for PL) (IL: Insertion loss, ER: Extinction ratio. * estimated)}.
\label{tab:comparison}
\end{table*}

\section{Discussion}

The generator model presented in this work demonstrates, in electromagnetic simulations, competitive broadband device performance within compact footprints across a broad spectrum of photonic functionalities. It also demonstrates utility across multiple fabrication platforms including EBL and PL. This approach provides multiple advantages compared to existing DRC-compliance methods in the literature. In contrast to building-block-based methods which ensure compliance by tiling the design region with fixed subwavelength primitives~\cite{KaiyuanWang2020}, our generator avoids reliance on rigid geometric definitions. This flexibility allows the optimization to operate within the fabrication-compatible design subspace without restricting geometry to discrete building blocks. As a result, it yields competitive performance and geometric diversity, without relying on discrete building blocks or resulting discretization artifacts. Compared with prior generator-based topology optimization approaches of two-phase projections~\cite{ChenHao2024} or brush-based painting~\cite{SchubertMartin2022}, our model directly learns a high-dimensional, fabrication-aware mapping. This architecture supports continuous design updates without violating DRC constraints. Consequently, it improves convergence stability and adapts more readily across disparate fabrication methodologies. Finally, in contrast to in-process regularized~\cite{ChenYuchen2023} or post-process regularized~\cite{HiesenerJacob2025} projection methods, as well as projection-filter approaches operating directly in the ambient device space~\cite{HammondAlec2021, RasmusChristiansen2021}, our generative framework encodes DRC-compliance as a structural property of the learned manifold. As this approach does not require externally scheduled parameters during device optimization, iterative tuning of regularization schedules is no longer necessary. By enforcing DRC compliance natively throughout the entire optimization, we eliminate the need for scheduled or iterative corrective steps such as the periodic projection-strength increments that raise the objective function and introduce artifacts or degrade performance in conventional methods. The combination of these advantages yields devices that consistently maintain smooth material boundaries while respecting minimum feature size constraints across a wide variety of photonic tasks.

An important feature of this proposed framework is that the generator and the 
electromagnetic simulation are fully decoupled. They are independent 
differentiable blocks connected only through gradient flow, with no shared 
resolution, discretization, or solver assumptions. The simulation block can 
therefore be swapped with any differentiable solver appropriate for the target 
device and fabrication platform, including 3D-FDFD, 3D-FDTD, or EME. 2D-FDFD 
simulations were used throughout this work for computational efficiency, 
following established practice in inverse-design methodology studies 
\cite{Hughes2018, AlexanderPiggott2015, PiggottAlexander2017, PiggottAlexander2020, Molesky2018, LoganSu2018}. Close agreement between inverse-designed devices and experimentally measured silicon photonic devices in the SOI platform and the same wavelength range has been established by Piggott 
et al.\ and Su et al.\ \cite{PiggottAlexander2017, AlexanderPiggott2015, LoganSu2018}. Since our framework 
is simulation-agnostic, this agreement transfers to our pipeline when coupled to 
a simulator of sufficient accuracy and spatial/temporal resolution for the target 
operation.

A quantitative comparison of simulation results in Table~\ref{tab:comparison} shows that our approach achieves competitive or lower simulated insertion loss across all three device categories within the evaluated configurations. The simulated insertion losses reported in Table 1 for the compared methods range from $0.24\,\text{dB}$ to $1.17\,\text{dB}$; within the same device type, our method achieves a simulated insertion loss as low as $0.03\,\text{dB}$ for the mode converter. While some of the compared designs achieve smaller footprints, this is accompanied by higher reported simulated insertion losses or a larger number of optimization iterations required. For instance, the structure transformation method reports $0.46\,\text{dB}$ simulated insertion loss in 450 iterations for the power splitter; our method converges to $0.06\,\text{dB}$ within 150 iterations on the same device type. For the wavelength demultiplexer, our design achieves a $16\,\text{dB}$ simulated extinction ratio and a simulated insertion loss of $0.25\,\text{dB}$, between those reported for TPPG and BBG in the loss-isolation trade-off. For the mode converter, our method yields a compact $2.1 \times 2.1\,\mu\text{m}^2$ footprint with a simulated insertion loss of $0.03\,\text{dB}$, in contrast to to the $0.64\,\text{dB}$ reported by the DRC-correction-based method within its published configuration.

The works compared in Table~1 operate under different minimum feature size 
constraints, ranging from ${\sim}70$~nm \cite{KaiyuanWang2020} to commercial foundry 
specifications \cite{HiesenerJacob2025}, which directly affects the accessible design 
topology and achievable insertion loss. While our EBL generator operates at 
60~nm (the finest constraint in this comparison), our PL generator operates at 
150~nm (comparable to \cite{ChenYuchen2023} and \cite{HiesenerJacob2025}). The performance difference 
between these two configurations, produced by the same method under different 
fabrication constraints, demonstrates the effect of minimum feature size on 
achievable insertion loss. A finer constraint provides access to a topological 
design space with greater degrees of design freedom. The relative performance of 
all entries in Table~1 reflects this relationship.

The comparisons presented in Table~1 draw from published results across 
different device footprints, simulation platforms, and optimization frameworks. 
Device performance is influenced by design area, discretization, solver accuracy, 
and convergence conditions. Because we impose DRC compliance during optimization, 
our devices utilize slightly larger footprints than some of the compared works, 
which may contribute to the lower simulated insertion losses. This may be 
attributed to the additional optimization degrees of freedom provided by the 
increased design area. Within these limitations, the results demonstrate that 
our latent-space reparameterization achieves competitive simulated insertion loss 
with computational efficiency advantages while maintaining intrinsic DRC compliance.

A direct comparison between the generator-based and pixel-based methods was also performed
for the wavelength duplexer on the 150 nm photolithography platform. The results are
presented in Supporting Information~11. In this case, the
generator-based method converges in 250 iterations, representing a 4.5-fold reduction
compared to the 1,116 iterations required by the pixel-based method. This is consistent
with the speedup observed for the 50/50 power splitter in Fig.~\ref{fig:figure3}. However, for this
PL-compatible duplexer, the pixel-based approach achieves a marginally better final loss
($5.4 \times 10^{-2}$) compared to the generator-based method ($8.7 \times 10^{-2}$),
yielding modestly improved spectral extinction ratios of 14--21 dB across the operational
bandwidth, compared to 10--14 dB for the generator-based design (see also Fig. S11 in
Supporting Information~11). This result indicates that in more constrained design scenarios
the pixel-based approach can reach a somewhat lower loss minimum given sufficient
computational budget. This is consistent with the expectation that the generative manifold
sacrifices some final device performance for a substantial reduction in the convergence
cost, by confining the search to a structured, learned, DRC-compliant subset. The
trade-off between convergence efficiency and final loss as well as the relative speedup are
consistent across both device types tested comparatively (splitter and duplexer). This
supports our earlier findings that the generator-based method characteristically yields
comparable device performance at a fraction of the optimization iterations.

Across the evaluated configurations, our approach achieves the lowest reported simulated insertion loss in each device category while maintaining competitive iteration counts and device footprints. These comparisons highlight the effectiveness and versatility of our generator across a diverse class of photonic devices. Although our presented framework achieves intrinsic DRC compliance across all 
device classes we examined, the current form of the topological loss function in 
Eq.~(3) enforces a spatially uniform minimum feature size. The erosion and 
dilation thresholds $\eta_e$ and $\eta_d$ are also global scalar constants 
applied identically at every pixel $(i,j)$. This formulation is suitable for 
single-etch planar waveguide devices governed by one foundry-specified minimum 
feature size. However, other classes of advanced nanophotonic structures may 
demand spatially heterogeneous length-scale constraints such as nanocavities with 
dielectric bowties \cite{Albrechtsen2022, Kountouris22}, or high-Q photonic 
crystals \cite{Christensen2022, Xiong24}. In these structures, subwavelength 
(${\sim}10$~nm) confining features coexist with surrounding structures governed 
by conventional foundry constraints (100--200~nm). Recent work has also 
demonstrated that this device class is moving toward atomic-scale realization 
through deterministic self-assembly \cite{Babar2023}, a regime fundamentally 
incompatible with a globally enforced minimum feature size. The presented 
generator framework can be extended to such geometries with a few modifications. 
First, the scalar thresholds in Eq.~\ref{eq:custom_loss} would be replaced by spatially indexed 
fields $\eta_e(i,j)$ and $\eta_d(i,j)$, guided by a zone mask that selectively 
defines constraints at the cavity center. Second, the generator architecture 
would be expanded into a multibranch design, in which a high-resolution branch 
generates fine cavity features and a coarse branch handles the surrounding 
structure. Additionally, the generator framework could serve the bulk of the 
design while a locally relaxed optimization stage refines the critical 
subwavelength zone, similar to a seeded topology optimization approach recently 
demonstrated \cite{HiesenerJacob2025}. These extensions constitute natural directions 
for future work that build upon the latent space reparameterization paradigm 
established here.

\section{Conclusion}

We have introduced a deep learning framework for silicon photonic inverse design that embeds fabrication constraints directly into the design representation. By mapping a low-dimensional latent space to intrinsically DRC-compliant geometries, our approach reformulates length-scale-constrained topology optimization as a scalable, unconstrained stochastic gradient problem. Unlike conventional projection-filter methods or building-block assembly, this framework unifies fabrication compatibility and optimization within a single differentiable generative model, confining the search entirely to the feasible design space at every step. Across multiple device classes including power splitters, wavelength duplexers, and mode converters, this method produces compact and high-performance components without requiring scheduled regularization or post-optimization refinement. When evaluated against previously reported results, our formulation achieves competitive insertion loss and footprints with a significantly reduced iteration count. All performance metrics reported in this work are based on electromagnetic simulations using finite-difference frequency-domain (FDFD) methods. The devices are intrinsically DRC compliant, and their experimental characterization remains a vital next step to confirm the simulated metrics. Ultimately, by treating fabrication constraints as a fundamental property of the design space rather than an external penalty, this work establishes a platform-agnostic pathway toward automated, intrinsically compliant nanophotonic design pipelines.

\section{Methods}

In both the conventional pixel-based optimization and our proposed generator-based approach, convergence is achieved as the objective successfully meets or drops below the pre-defined target. This convergence condition ensures that the optimization process continues iteratively until the difference between the current device performance and the target goal is eliminated, signaling that the design requirements have been satisfied. For training both the EBL generator model and the PL generator model, we used the ADAM optimizer with a learning rate of $1 \times 10^{-3}$, $\beta_1 = 0.9$, and $\beta_2 = 0.999$, using a batch size of 16. The training was performed for 2060 iterations for the electron-beam generator and 1500 iterations for the PL generator, during which the slope of the scheduled softmax layer was gradually increased from 1 to $10^{30}$, effectively approximating a step function by the end of training. All training processes were executed on an NVIDIA A100 GPU using the Flax~\cite{HeekJ2024} and JAX~\cite{BradburyJ2018} libraries, with Optax~\cite{DeepMind2020} managing the optimization. For the conventional optimization of the 50/50 power splitter, we used the method of moving asymptotes (MMA) optimization algorithm from the NLopt package~\cite{JohnsonNLopt}, which proves to be the most effective in the context of pixel-based optimization in our implementation. Automatic DRC verification of the generated and optimized devices was performed using KLayout. The FDFD solver employed a uniform grid discretization of 25 nm × 25 nm throughout all optimizations, providing adequate spatial resolution for capturing electromagnetic field variations while maintaining computational efficiency for iterative design. Since the electromagnetic solver (FDFD) constituted the dominant computational cost compared with the generator inference, we employed factorization caching via the Intel oneMKL PARDISO library~\cite{IntelPARDISO2023} and executed all optimization steps on a $2.4\,\text{GHz}$ Intel Xeon Gold processor with eight cores.

\begin{acknowledgments}
This work is supported by the Scientific and Technological Research Council of Turkey (TUBITAK) under grant number 122E214.
\end{acknowledgments}

\section*{DATA AVAILABILITY}
The data that support the findings within this manuscript are available from the corresponding author upon reasonable request.

\section*{CODE AVAILABILITY}
The trained model architectures and corresponding weights for both the EBL and PL design frameworks are publicly available at \url{https://github.com/Photonic-Architecture-Laboratories/drcgenerator}.

\section*{COMPETING INTEREST}
The authors declare no competing interests.

\def\bibsection{\section*{References}}
\bibliographystyle{naturemag}
\bibliography{main}

\end{document}


\raggedbottom
\preprint{APS/123-QED}

\title{Intrinsically Design-Rule-Compliant Nanophotonic Inverse Design via Learned Generative Manifolds: \\ Supporting Information}

\author{Bahrem~Serhat~Danis}
\affiliation{\mbox{Dept. of Electrical and Electronics Engineering, Koç University, Istanbul, 34450, Turkey}}

\author{Demet~Baldan~Desdemir}
\affiliation{\mbox{Dept. of Electrical and Electronics Engineering, Koç University, Istanbul, 34450, Turkey}}

\author{Enes~Akcakoca}
\affiliation{\mbox{Dept. of Electrical and Electronics Engineering, Koç University, Istanbul, 34450, Turkey}}

\author{Zeynep~Ipek~Yanmaz}
\affiliation{\mbox{Dept. of Electrical and Electronics Engineering, Koç University, Istanbul, 34450, Turkey}}

\author{Gulzade~Polat}
\affiliation{\mbox{Dept. of Electrical and Electronics Engineering, Koç University, Istanbul, 34450, Turkey}}

\author{Ahmet~Onur~Dasdemir}
\affiliation{\mbox{Dept. of Electrical and Electronics Engineering, Koç University, Istanbul, 34450, Turkey}}

\author{Aytug~Aydogan}
\affiliation{\mbox{KU Leuven, Dept. of Physics and Astronomy, B-3000 Leuven, Belgium}}

\author{Abdullah~Magden}
\affiliation{\mbox{Dept. of Mathematics, Faculty of Engineering and Natural Sciences, Bursa Technical University, Bursa, 16310, Turkey}}

\author{Emir~Salih~Magden}
\email{Corresponding author: esmagden@ku.edu.tr}
\affiliation{\mbox{Dept. of Electrical and Electronics Engineering, Koç University, Istanbul, 34450, Turkey}}
\affiliation{\mbox{KUIS AI, Koç University, Istanbul, 34450, Turkey}}


\maketitle


\section{Topological Loss Function}

The topological loss term~\cite{HammondAlec2021} is designed as a combination of two distinct design-rule constraints corresponding to the minimum linewidth and minimum line spacing. It serves to enforce manufacturability constraints during the optimization process by penalizing structural features that violate these design rules
\begin{equation}
\mathcal{L}
=
\sum_{i,j}
I^{\mathrm{LW}}_{i,j}(\rho)\,
\left[\min\!\left(\rho_{i,j} - \eta_e,\, 0\right)\right]^2
+
I^{\mathrm{LS}}_{i,j}(\rho)\,
\left[\min\!\left(\eta_d - \rho_{i,j},\, 0\right)\right]^2 .
\label{eq:topological_loss}
\end{equation}
Here, $\rho$ denotes the normalized permittivity profile ranging from 0 ($\text{SiO}_2$) to 1 (Si). The parameters $\eta_d$ and $\eta_e$ act as threshold values that govern the dilation and erosion steps, respectively. Two auxiliary indicator functions, $I_{\text{LW}}$ and $I_{\text{LS}}$, are then constructed to identify the silicon and silica domains based on these thresholded regions.
\begin{equation}
I^{\mathrm{LW}}(\rho, \tilde{\rho}) = \tilde{\rho} \, \exp\!\left(-\alpha \lvert \nabla \rho \rvert^{2}\right)
\label{eq:I_LW}
\end{equation}
\begin{equation}
I^{\mathrm{LS}}(\rho, \tilde{\rho}) = \left(1 - \tilde{\rho}\right) \, \exp\!\left(-\alpha \lvert \nabla \rho \rvert^{2}\right)
\label{eq:I_LS}
\end{equation}
In this formulation, $\tilde{\rho}$ represents the generator output after applying a 2D Gaussian blur with radius $R$, while $\alpha$ controls the relative contribution of the two auxiliary indicators. The threshold parameters $\eta_d$ and $\eta_e$ are linked to the minimum linewidth $l_w$, minimum line spacing $l_s$, and the Gaussian filter radius, since violations of these design rules are detected through the filtering process. As the filter radius determines the effective smoothing scale, its relation to $l_w$ and $l_s$ dictates how these thresholds are selected. Therefore, $\eta_d$ and $\eta_e$ intrinsically depend on $l_w$, $l_s$, and $R$. During training, the topological loss incorporates indicator functions that identify inflection regions and exploit the known relationship between the filter kernel and the morphological transformations, specifically the erode and dilate operations, thereby guiding the generator toward designs that rigorously satisfy geometric constraints.
\newpage
The threshold parameters $\eta_d$ and $\eta_e$ are determined directly from the foundry-specified minimum linewidth ($l_w$) and minimum spacing ($l_s$), following the methodology outlined in previous studies~\cite{HammondAlec2021,QianXiaoping2013}. These studies emphasized that maintaining strict control over the smallest permissible feature widths and gap distances is essential for preventing numerical artifacts and ensuring smooth, fabrication-consistent topologies in the optimization process:

\begin{equation}
\eta_e =
\begin{cases}
\dfrac{1}{4}\left(\dfrac{l_w}{R}\right)^2 + \dfrac{1}{2}, 
& \dfrac{l_w}{R} \in [0,1], \\[6pt]
-\dfrac{1}{4}\left(\dfrac{l_w}{R}\right)^2 + \dfrac{l_w}{R}, 
& \dfrac{l_w}{R} \in [1,2], \\[6pt]
1, 
& \dfrac{l_w}{R} \in [2,\infty).
\end{cases}
\label{eq:eta_e}
\end{equation}

\begin{equation}
\eta_d =
\begin{cases}
\dfrac{1}{2} - \dfrac{1}{4}\left(\dfrac{l_s}{R}\right)^2, 
& \dfrac{l_s}{R} \in [0,1], \\[6pt]
1 + \dfrac{1}{4}\left(\dfrac{l_s}{R}\right)^2 - \dfrac{l_s}{R}, 
& \dfrac{l_s}{R} \in [1,2], \\[6pt]
0, 
& \dfrac{l_s}{R} \in [2,\infty).
\end{cases}
\label{eq:eta_d}
\end{equation}

These specific values are determined from the ratio of the minimum linewidth to the minimum spacing, ensuring accurate thresholding, allowing the correct identification of regions that violate the design rules for linewidth and line spacing. Following the methodology in previous studies\cite{HammondAlec2021, ZhouMingdong2015, Christiansen2020}, we use $l_s = l_w$ and set the Gaussian filter radius to twice the minimum feature size, $R = 2l_w = 2l_s$, which corresponds to typical values reported in the literature. Under these conditions, the threshold parameters are assigned as $\eta_e = 0.75$ and $\eta_d = 0.25$. In our generator implementation, the minimum linewidth and spacing are explicitly defined according to the fabrication process. For the standard electron-beam lithography (EBL) generator model, the minimum feature size is set to $60\,\text{nm}$, resulting in a Gaussian filter radius of $120\,\text{nm}$. For the photolithography (PL) generator model, the minimum feature size is $150\,\text{nm}$, with a corresponding filter radius of $300\,\text{nm}$. This configuration ensures that the topological loss accurately reflects process-specific fabrication limits while maintaining consistent morphological control. In addition, the damping coefficient $\alpha$ is defined as the fourth power of the grid resolution, $\alpha = (1/25\,\text{nm})^4 = 2.56 \times 10^{-6}\,\text{nm}^{-4}$. This value is sufficient for these hyperparameters to accurately detect regions where the linewidth constraints are violated.

\newpage
\section{Boundary Design Rule Compliance and Waveguide Integration}

\begin{figure*}[ht!]
    \centering
    \includegraphics[width=\textwidth]{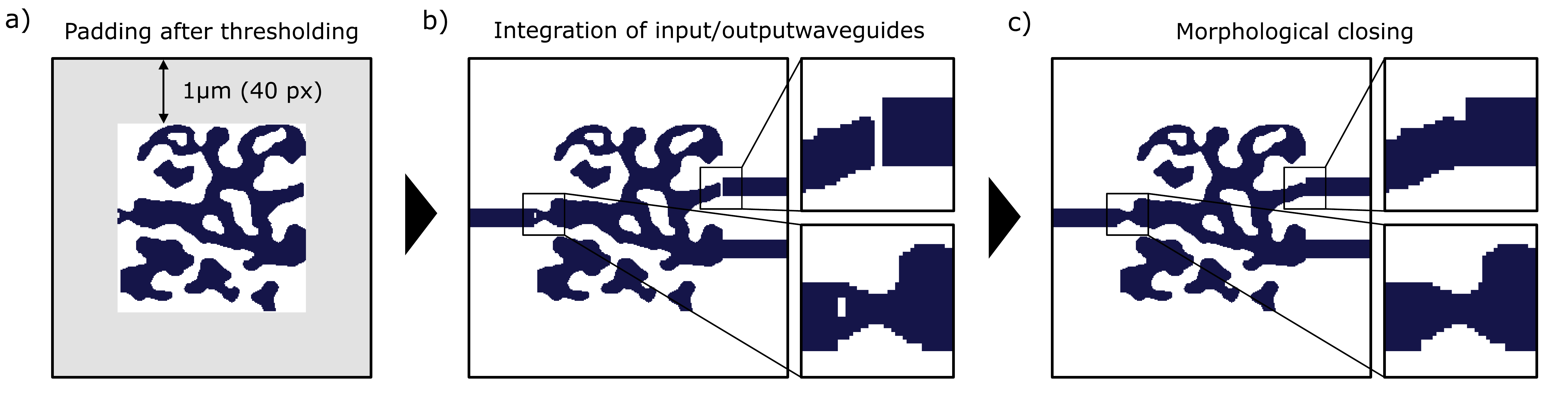}
    \caption{Step-by-step processing pipeline for boundary DRC compliance and GDS layout 
generation. (a) SiO$_2$ padding applied to all edges of the thresholded device. 
(b) Integration of input/output waveguides into the padded binary layout. 
(c) Morphological closing applied to the combined structure, resolving all 
boundary and waveguide-junction DRC violations.}
    \label{fig:sfigure1}
\end{figure*}

During training of the generators, after thresholding, every generated 
permittivity distribution is padded with a $1~\mu$m (40-pixel) SiO$_2$ region 
along all four edges. 450-nm-wide input and output waveguides are added along 
the left and right edges, with randomized count and lateral positions per sample. 
This is to ensure the architecture generalizes well to devices with multiple 
input and/or output waveguides. Morphological closing is then applied to the 
padded, waveguide-attached structure before the DRC loss is evaluated. This 
procedure explicitly exposes the device boundaries and both waveguide--device 
connections, so the generator learns to produce distributions that remain 
DRC-compliant at these potentially problematic regions. Crucially, the closing 
kernel is sized slightly above the minimum feature size, so that any sub-length-scale gaps or islands at the device 
boundaries or waveguide connections are filled or merged. These kernel footprints ($75~\text{nm} \times 75~\text{nm}$ for EBL and $175~\text{nm}
\times 175~\text{nm}$ for PL) therefore serve as practical lower bounds on both the
minimum area and minimum enclosed area for the respective generator configurations. During inference 
(inverse design), a similar procedure is applied. Input and output waveguides 
are appended to the binary layout after thresholding. Morphological closing is 
again applied to the complete structure, so that no new violations appear at the 
boundaries or waveguide connections. This procedure is illustrated in 
Fig.~\ref{fig:sfigure1}.

\section{Training Latent Space Dataset with Spatial Feature Diversity}

To ensure that the training dataset encompasses samples capturing both coarse and fine spatial features, we generated 1,000 unique noise patterns. This diverse dataset is crucial for minimizing overfitting and maximizing the model's generalizability across varied latent space inputs. Mathematically, each training sample $I(x,y)$ is constructed as a superposition of $D$ independent noise components, $\eta_d$, with each modulated by distinct scale ($s$) and offset ($O_x, O_y$) parameters.

\begin{equation}
\label{eq:intensity_formula}
I(x, y) = \frac{1}{D} \sum_{d=1}^{D} \eta_d \left( \frac{x + O_x}{s}, \frac{y + O_y}{s} \right),
\end{equation}

This construction allows us to synthesize a broad spectrum of structural complexity, thoroughly exploring the feature space of the latent representation. The core of this synthesis lies in the scale parameter $s$, which directly controls the frequency content of the noise: larger values generate low-frequency, coarse features, while smaller values produce high-frequency, fine details. Concurrently, the offset parameters $O_x$ and $O_y$ introduce translational shifts to the pattern, effectively sampling different phase spaces of the noise field. By averaging over $D$ independent fields, we achieve a statistically robust and richer combination of features, far surpassing simple random sampling. The $d$-th stochastic noise field, $\eta_d(u,v)$, is defined by a pseudo-random gradient interpolation function, which ensures the generated noise possesses perceptually coherent, continuous structures rather than being uniform white noise. The field is mathematically defined using a summation over grid points $(i, j)$:

\begin{equation}
\label{eq:basis_function}
\eta_d(u,v) = \sum_{i=0}^{1} \sum_{j=0}^{1} w_i(u) w_j(v) \mathbf{g}_{d,ij} \cdot (u - i, v - j).
\end{equation}

Here, $\mathbf{g}_{d,ij}$ are pseudo-random gradient vectors assigned at each lattice point $(i,j)$; $w_i(u)$ is the smooth fade (interpolation) function; $(x,y)$ are pixel coordinates in the 2D spatial domain; $s$ is the scale controlling spatial frequency; $(O_x, O_y)$ are spatial offsets introducing positional shifts; and $D$ is the number of independent noise channels controlling structural complexity. Finally, each noise map is normalized to the unit interval as:

\begin{equation}
\label{eq:normalization}
I'(x, y) = \frac{I(x, y) - \min(I)}{\max(I) - \min(I)}.
\end{equation}

This formulation explicitly defines the generation process used by the \texttt{noise} library\cite{Caseman2015}, ensuring reproducible construction of noise fields with tunable spatial diversity across multiple scales and offsets. We generated 1,000 samples by systematically varying the scale, offset, and noise dimensionality. These variations produced multiple granularities across different topologies, thereby enhancing the generalization capabilities of the generator. Specifically, the scale parameter modulates the spatial frequency of the noise, controlling whether the resulting patterns exhibit fine or coarse textures. The offset introduces spatial shifts, allowing the generator to encounter diverse spatial alignments. The noise dimensionality defines the number of independent stochastic components shaping each pattern, thereby modulating structural complexity and diversity.

\begin{figure*}[ht!]
    \centering
    \includegraphics[width=\textwidth]{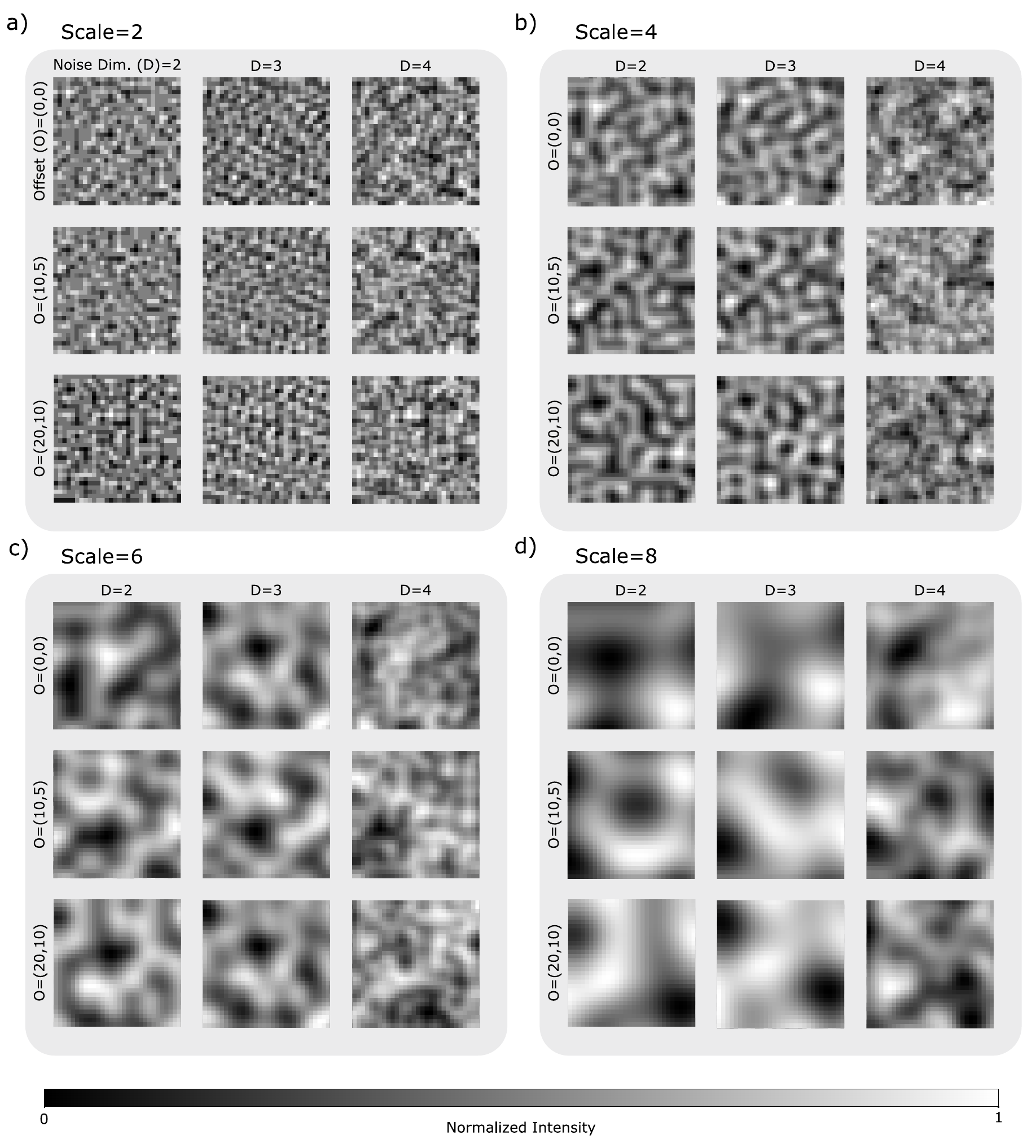}
    \caption{Effect of noise hyperparameters on spatial features. Two-dimensional noise maps (30 $\times$ 30 pixels) were generated by varying the scale, offset (O), and noise dimensionality (D). (a) Scale=2, (b) Scale=4, (c) Scale=6, (d) Scale=8, with O = (0, 0), (10, 5), (20, 10) (rows) and D = 2, 3, 4 (columns). Increasing the scale smooths the spatial features, while higher dimensionality introduces finer variations. These maps represent representative latent-space samples used for training the electron-beam lithography and photolithography generator models. Grayscale denotes normalized intensity from 0 (black) to 1 (white).}
    \label{fig:sfigure2}
\end{figure*}

To show the effect of noise hyperparameters on spatial features, we varied the scale, offset, and dimensionality for a fixed $30 \times 30$ 2D noise map. As can be seen from Fig.~\ref{fig:sfigure2}, the scale takes values (a) 2, (b) 4, (c) 6, and (d) 8; offsets are set as integer pairs such as $(0, 0)$, $(10, 5)$, and $(20, 10)$; and the noise dimensionality is 2, 3, or 4. This figure illustrates how spatial features change under fixed input conditions.
\begin{figure*}[ht!]
    \centering
    \includegraphics[width=\textwidth]{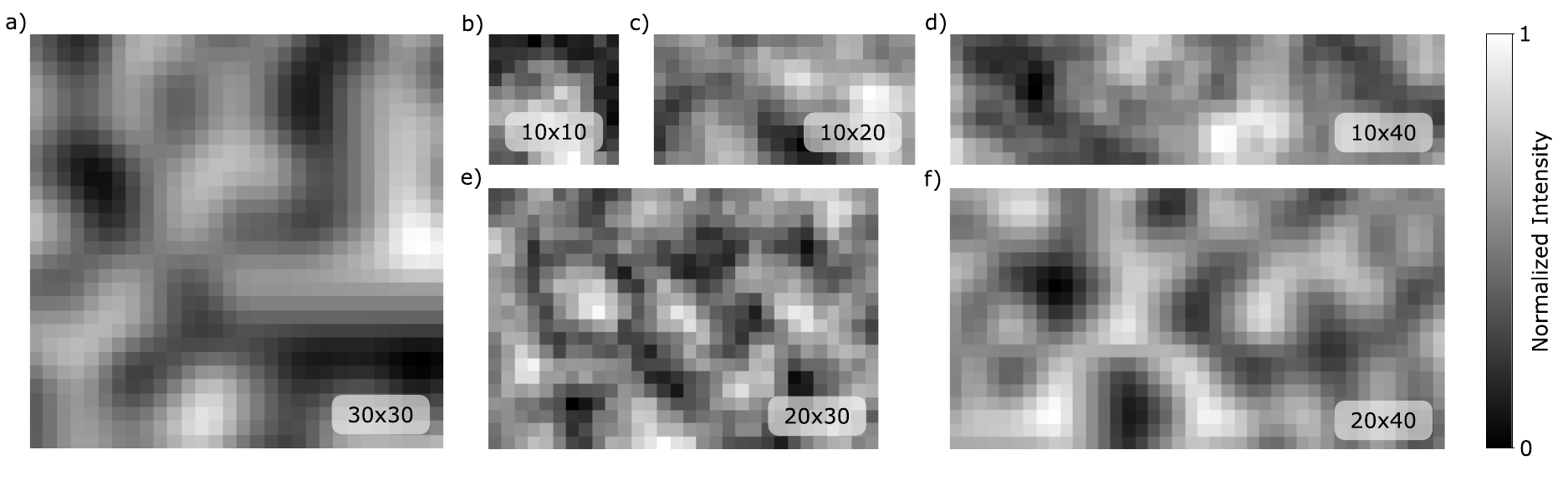}
    \caption{Representative training samples of varying sizes generated under different noise hyperparameter settings: (a) $30 \times 30$, (b) $10 \times 10$, (c) $10 \times 20$, (d) $10 \times 40$, (e) $20 \times 30$, and (f) $20 \times 40$ pixels. The multi-scale latent space realizations illustrate how variations in latent space size and noise hyperparameters shape the normalized intensity distributions.}
    \label{fig:sfigure3}
\end{figure*}

During sample generation, the scale parameter was randomly selected as an integer between 2 and 16. Offsets were chosen as random integer pairs, with minimum values of 0 and maximum values determined by the corresponding sample dimensionality. The noise dimensionality was randomly selected as 2, 3, or 4 for each sample. Additionally, to enable the training of a size-agnostic generator capable of producing layouts of various dimensions, the sample size was randomly assigned as an integer between 4 and 45, independently for both $x$ and $y$ dimensions. Samples generated from the training latent space dataset, each using randomly picked noise hyperparameters, can be seen in Fig.~\ref{fig:sfigure3}. This figure specifically illustrates the variety of sample dimensions (e.g., (a) $30 \times 30$, (b) $10 \times 10$, (c) $10 \times 20$, (d) $10 \times 40$, (e) $20 \times 30$, and (f) $20 \times 40$ pixels) used in the training dataset to achieve the size-agnostic capability. This extensive variation in the input dimensions ensures the model learns the underlying latent relationships independently of the fixed output size. Because the generator learns local, scale-invariant transformations (instead of device-specific geometries), the fully convolutional architecture allows size-agnostic inference at any footprint, including sizes not encountered during training. Consequently, the generator is robustly trained across a wide range of dimensionality, allowing it to generalize effectively when generating patterns for arbitrary $x$ and $y$ dimensions not seen during training.

\section{Input-Output Size Mapping in the Electron-Beam and Photolithography Generator Models}

The input space of the model consists of grayscale two-dimensional distributions with pixel intensities normalized within the range $[0, 1]$. The minimum input dimension is $4 \times 4$, while the maximum size can be arbitrarily chosen depending on the desired spatial scale. In our training, the grayscale latent-space samples were generated up to $60 \times 60$ pixels. Since both generator models are composed entirely of convolutional layers followed by pointwise nonlinear functions, they are inherently size-agnostic. This means that the same model architecture can process inputs of arbitrary spatial dimensions, effectively allowing the generation of physically larger patterns without structural modification. The mapping between the input and output sizes exhibits an approximately linear relationship; each input pixel corresponds to a predictable number of output pixels, enabling one-dimensional interpolation of the scaling behavior. For the EBL generator, the effective scaling factor is approximately 11, while for the PL generator, it is around 18. However, due to discrete convolution operations involving floor rounding, the mapping is not perfectly linear.

\begin{figure*}[ht!]
    \centering
    \includegraphics[width=\textwidth]{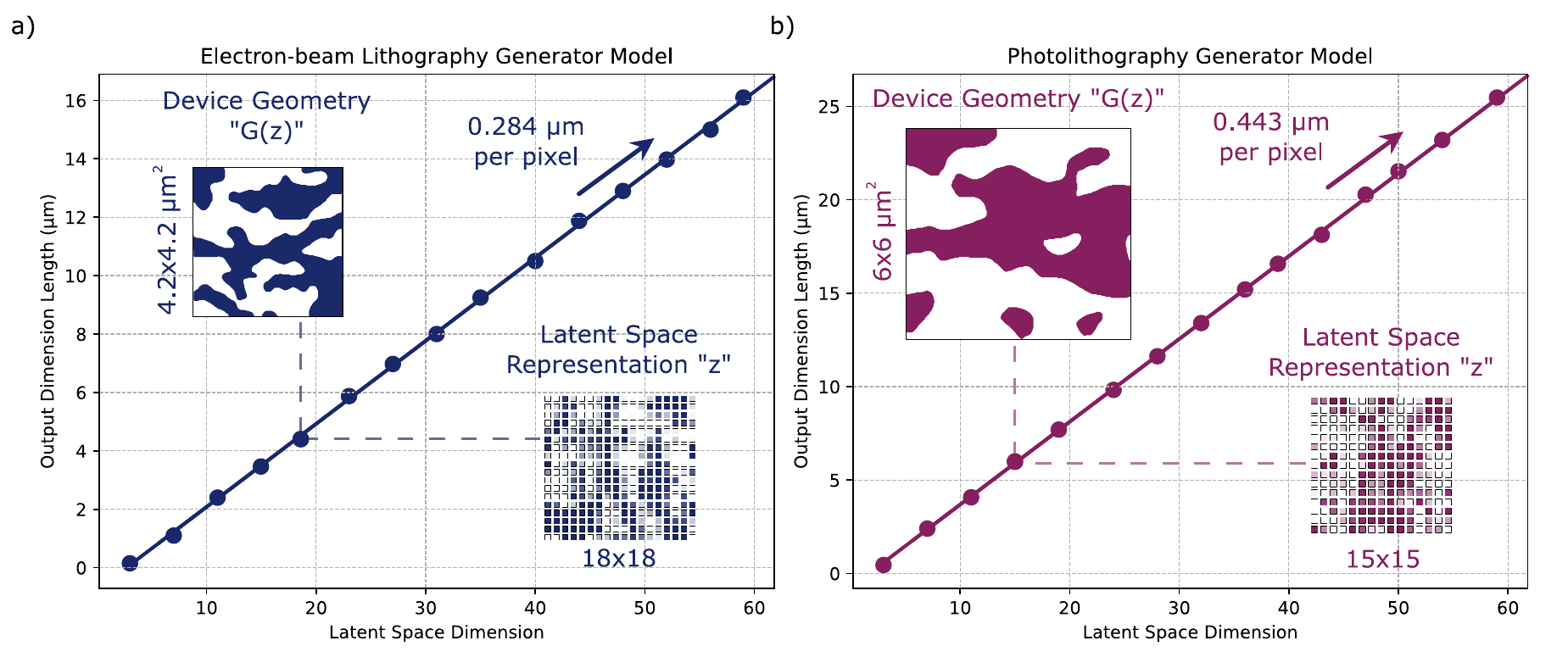}
    \caption{Latent-space interpolation and physical mapping for EBL and PL generators. (a) EBL generator and (b) PL generator: interpolation of latent-space dimensions to physical output dimensions. Insets: top-left, representative latent-space vector; bottom-right, corresponding output structure from the trained generator, with indicated physical dimensions. The comparison highlights the distinct spatial resolutions: the EBL generator produces high-resolution outputs ($\sim$0.284~$\mu$m/pixel equivalent), whereas the PL generator yields coarser patterns ($\sim$0.443~$\mu$m/pixel).}
    \label{fig:sfigure4}
\end{figure*}

Both generator outputs are defined on a geometrical grid resolution of $25\,\text{nm}$ in the space of the device geometry. Also, the electromagnetic simulations performed during both training and optimization 
uses the same $25~\text{nm} \times 25~\text{nm}$ grid discretization in the 
FDFD solver for computational consistency. However, the generator output can be 
resampled to any simulation grid via interpolation either during or after 
optimization. As shown in Fig.~\ref{fig:sfigure4}, after interpolation, a single latent-space pixel represents approximately $30\,\text{nm}$ in the device space of the EBL generator and $50\,\text{nm}$ in that of the PL generator. This difference highlights the distinct spatial resolutions of the two models: the EBL generator produces high-resolution patterns ($\approx 0.284\,\mu\text{m}/\text{pixel}$ equivalent), whereas the PL generator operates at a lower resolution ($\approx 0.443\,\mu\text{m}/\text{pixel}$). Fig.~\ref{fig:sfigure4} insets illustrate the translation of a shared latent-space input into the respective physical domains of each generator. Each model produces geometries tailored to its specific fabrication regime, ensuring that the resulting structures satisfy the inherent resolution limits and design rules of either EBL or PL.

\section{Discreteness Metric $\Gamma$ for Monitoring Binarization and Projection Strength Parameter $\beta$}

To quantitatively assess how well the material boundaries between $\text{SiO}_2$ and $\text{Si}$ are defined in the design region, we define a discreteness metric ($\Gamma$) that measures how close each design pixel (or voxel) is to the two binary material states (0 for $\text{SiO}_2$, and 1 for $\text{Si}$). This metric provides an interpretable scalar measure of how closely the density distribution $\rho(x, y) \in [0, 1]$ approaches a binary state. Here, $\Gamma = 1$ corresponds to a completely binarized structure and $\Gamma = 0$ represents a fully grayscale distribution. While the metric is applicable to any two-material system, the values $\rho(x, y) = 0$ and $\rho(x, y) = 1$ in this study specifically represent the $\text{SiO}_2$ and $\text{Si}$ phases. Intermediate values denote partially mixed or non-binarized regions.

To compute the discreteness, the design region is first divided into $N_x \times N_y$ subregions denoted by $\Omega_{i,j}$, where $i = 1, 2, \dots, N_x$ and $j = 1, 2, \dots, N_y$. The subregion division ensures that the metric captures spatial variations in binarization across the entire structure. For each subregion $\Omega_{i,j}$, the local discreteness value $\Gamma_{i,j}$ is defined as the ratio of pixels whose density values fall within a discrete tolerance range near 0 or 1, relative to the total number of pixels in that subregion:
\begin{equation} 
\Gamma_{i,j} = \frac{N_{\text{discrete}}^{(i,j)}}{N_{\text{total}}^{(i,j)}} \end{equation}
where
\begin{equation}
N_{\text{discrete}}^{(i,j)} = \# \left\{ \rho(x,y) \in \Omega_{i,j} \mid \rho(x,y) \leq \tau \text{ or } \rho(x,y) \geq 1 - \tau \right\}.
\end{equation}
Here, $\tau$ is the discreteness tolerance parameter (typically set between 0.01 and 0.05) that determines how close a pixel's value must be to either binary state to be considered discrete. For the pixel-based optimization performed in the main text, we set $\tau = 0.01$. $N_{i,j}^{\text{total}} = |\Omega_{i,j}|$ denotes the total number of pixels in the subregion. The overall discreteness of the structure is obtained by averaging the local values over all subregions:
\begin{equation} 
\Gamma = \frac{1}{N_x N_y} \sum_{i=1}^{N_x} \sum_{j=1}^{N_y} D_{i,j} 
\end{equation}
A higher value of $\Gamma$ indicates that most of the design pixels have converged toward binary values, meaning the device is closer to a manufacturable topology with well-defined material boundaries. Conversely, a lower $\Gamma$ signifies that a significant portion of the structure remains in an intermediate, non-binary state.

For the conventional optimization, we initialize the projection strength parameter $\beta$ at 8. As illustrated in Fig.~\ref{fig:sfigure1}(a), whenever the optimization does not achieve sufficient improvement, we double the value of $\beta$, leading to an exponential increase. This strategy guarantees that the final design is fully binarized. The loss curve reveals that each increase in $\beta$ initially raises the loss due to sudden changes in design parameters, while simultaneously improving the discreteness metric. The discreteness metric $\Gamma$ quantifies the degree to which the design is binary or grayscale. Therefore, as $\beta$ increases, the minimum achievable loss also worsens, indicating that the optimization problem becomes progressively more challenging.

\section{Training the Photolithography Generator Model}

The PL generator model differs from the EBL generator model in two principal aspects. First, the overall upsampling ratio in the cascaded upsampling stages was increased from 1.4 to 1.6, resulting in a total latent-space scaling factor change from $1.4^4$ to $1.6^4$. Second, the convolutional architecture was streamlined: the number of convolutional layers was reduced from three to two, while the kernel size was enlarged from $5 \times 5$ to $7 \times 7$. This modification allows the network to better enforce design-rule constraints at larger feature scales. Additionally, the kernel size of the final morphological closing operation was expanded from $3 \times 3$ to $7 \times 7$ to ensure smoother feature boundaries and improved mask continuity.

\begin{figure*}[ht!]
    \centering
    \includegraphics[width=0.86\textwidth]{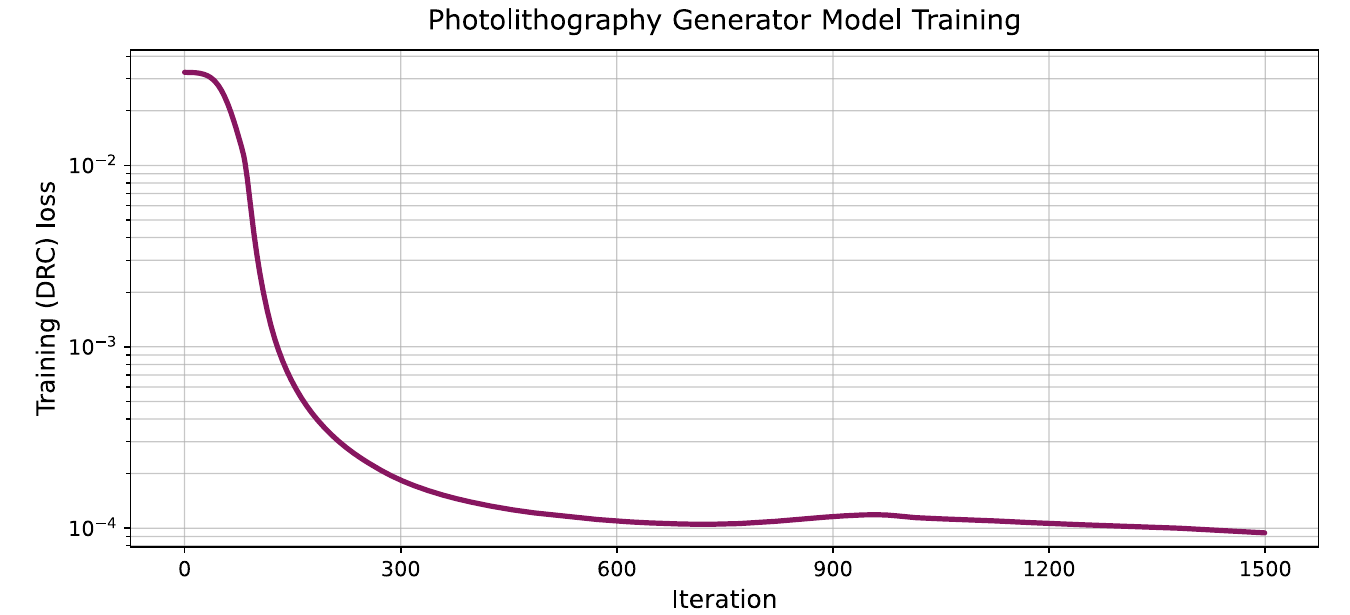}
    \caption{Training dynamics of the photolithography generator. The design rule check loss decreases from $\sim 3 \times 10^{-2}$ to $\sim 1 \times 10^{-4}$ over 1500 iterations, showing a rapid reduction in the first 300 iterations and a gradual approach to saturation, indicating stable convergence of the model.}
    \label{fig:sfigure5}
\end{figure*}

For training, we employed the same dataset used in the EBL generator model, ensuring a consistent comparison. Optimization was performed using the Adam optimizer, with hyperparameters identical to those reported in the Methods section. The training loss evolution for the PL generator is shown in Supplementary Fig.~\ref{fig:sfigure5}, where the design rule check (DRC) loss begins at approximately $3 \times 10^{-2}$ and rapidly decreases during the first 300 iterations, reaching a near-saturation regime. By around 1500 iterations, the DRC loss converges to approximately $10^{-4}$. This progressive reduction correlates with a gradual decrease in the number of DRC violations across generated device layouts. A DRC loss value on the order of $10^{-4}$ indicates a well-constrained design region with a minimum feature size of about $150\,\text{nm}$. Compared to the electron-beam generator, which achieves a minimum DRC loss of roughly $8 \times 10^{-6}$, the slightly higher final loss in the PL model originates from the larger Gaussian filter radius used during training. Increasing the filter size effectively enlarges the minimum allowable feature size in the topology, thereby increasing the norm of the gradients within the indicator functions. Consequently, the achievable minimum loss becomes bounded from below by this feature-size scaling. Moreover, the total number of trainable parameters in the PL generator was substantially reduced, from 13,243 in the electron-beam model to 1,619, leading to a shorter training duration of 1,500 iterations (compared to 2,060 iterations in the electron-beam case). This reduction accelerates convergence while preserving sufficient model expressivity for mask generation under photolithographic constraints.

\section{Statistical Validation of DRC-Compliant Generation Across the Latent Space}

\begin{figure*}[ht!]
    \centering
    \includegraphics[width=\textwidth]{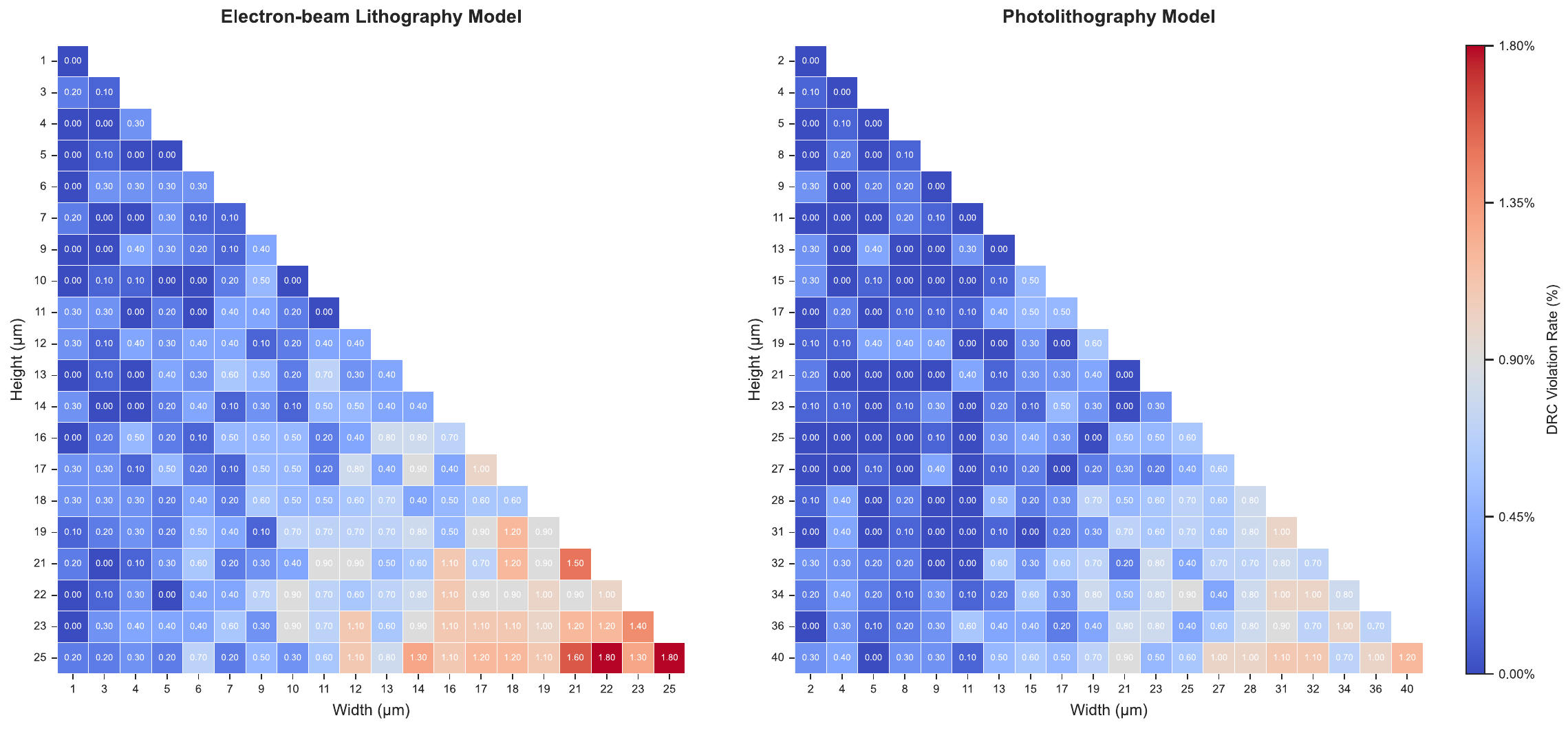}
    \caption{DRC violation rate (\%) as a function of device footprint for the EBL generator 
(left) and the PL generator (right), evaluated over 1,000 randomly sampled 
layouts per (width, length) pair using automated design rule checking in KLayout. 
The violation rate increases gradually with device area but remains below $1.8\%$ 
for the EBL model and $1.2\%$ for the PL model.}
    \label{fig:sfigure6}
\end{figure*}

For a strict guarantee of DRC compliance, the learned manifold $G(\mathcal{Z})$ 
must entirely lie within the fabrication-compliant subset $\mathcal{S}$. This 
would require the topological training loss to reach its global minimum over very 
large training sets. However, for SGD-trained generative networks, no such 
analytically provable guarantee exists. Instead, what we can provide is empirical 
evidence that the learned manifold is (for practical purposes) contained within 
$\mathcal{S}$. In Fig.~\ref{fig:sfigure6}, we plot DRC results obtained from KLayout for 1,000 
independent devices generated from random latent-space distributions using the 
same Perlin noise parameter distributions employed during our generator training. 
This procedure has been performed both for our e-beam and photolithography 
generators, for latent dimensions ranging from 5 to 90 pixels on both the width 
and the length axes.

Within the device footprint range used in this work (up to $7 \times 7~\mu$m$^2$), 
the violation rate is identically zero for both generators across all samples 
tested. Beyond this range, rates increase gradually, reaching $1.8\%$ for the 
EBL model at $25 \times 25~\mu$m$^2$ and $1.2\%$ for the PL model at 
$40 \times 40~\mu$m$^2$ (which are both well beyond any device shown in our 
example demonstrations). The PL generator exhibits consistently lower rates due 
to its larger convolutional kernels and the correspondingly coarser 150~nm 
minimum feature size. These results confirm that while there is no strict 
guarantee, the learned manifold is almost entirely contained within $\mathcal{S}$.

\section{Devices Designed with the Photolithography Model and Their Optimization Details}

Using the PL-based generator model, we optimized the same three representative classes of photonic devices: power splitters, a photonic duplexer, and a mode converter. The simulation parameters, including the wavelength range, the $\text{Si}/\text{SiO}_2$ material platform, input light polarization, and the input–output waveguide widths, are kept identical to those used in the EBL configuration to ensure a fair comparison. Following optimization, each device design is subjected to an automated DRC, and all designs successfully passed without any violations. To translate these optimized designs into physically realizable patterns, the generator's inference stage plays a crucial role. During the inference of the trained generator, the scheduled softmax is replaced with a thresholding operation to obtain a binary layout. Since this thresholding layer is non-differentiable, its gradient is approximated using the straight-through estimator introduced earlier, allowing the entire generator and pipeline to remain differentiable and enabling gradient-based optimization of the latent space. The results and performance analyses of these three device categories are reported in the following sections.

\subsection{Broadband Power Splitters with Variable Splitting Ratios}

Following the same design strategy as in the EBL model, three distinct power splitters are designed, each targeting a different splitting ratio. For all designs, the optimization process is performed using six discrete wavelength sampling points. The desired output power distributions are set to 50/50, 30/70, and 10/90 between the two output waveguides across the sampled wavelengths. Each optimized power splitter occupies a compact footprint of $6 \times 6\,\mu\text{m}^2$. For all the photolithography-based power splitters, the 
output waveguide separation is designed to be $2.5~\mu$m, measured center to 
center between the two output waveguides.

\begin{figure*}[ht!]
    \centering
    \includegraphics[width=\textwidth]{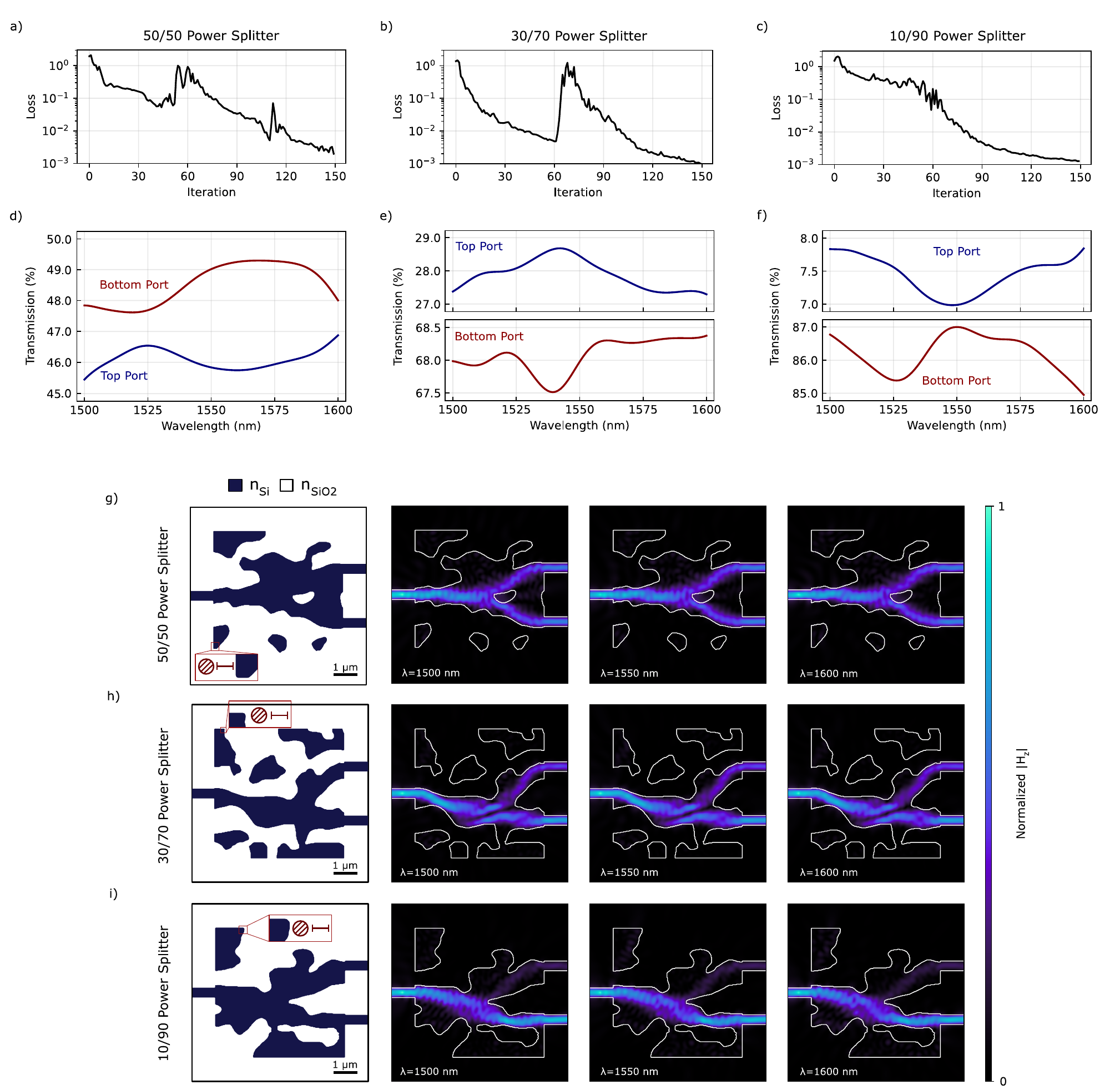}
    \caption{Ultrabroadband power splitter performance for different splitting ratios designed with the photolithography generator model. (a)--(c) Evolution of the optimization loss during the training of the photolithography generator model for 50/50 (a), 30/70 (b), and 10/90 (c) power splitters, illustrating rapid reduction in loss across 150 iterations. (d)--(f) Spectral transmission for the top and bottom ports of the respective devices, showing consistent splitting behavior over the 1500--1600~nm wavelength range. (g)--(i) Left panels depict the final permittivity maps (silicon in dark blue, SiO$_2$ in white), while the right panels show the corresponding normalized magnetic field ($H_z$) distributions at $\lambda = 1500$, 1550, and 1600~nm. The field profiles confirm ultrabroadband operation across all splitter designs.}
    \label{fig:sfigure7}
\end{figure*}

For the 50/50 power splitter (Fig.~\ref{fig:sfigure7}(a)), the optimization process rapidly reduces the loss from 1.9 to $2 \times 10^{-3}$ within 150 iterations, resulting in a well-balanced power distribution between the two output ports. As shown in Fig.~\ref{fig:sfigure7}(d), both the generator-based and conventional designs maintain transmission levels of 45--50\% at each port over the wavelength range of 1,500--1,600\,nm. The optimized splitter exhibits an overall transmission of 97\%, corresponding to an insertion loss of $0.22\,\text{dB}$, demonstrating highly efficient power division. For the 30/70 power splitter, the optimization was performed using five sampling points, targeting 30\% transmission at the top port and 70\% at the bottom port. As shown in Fig.~\ref{fig:sfigure7}(b), the loss decreases from 1.38 to $1 \times 10^{-3}$ after 150 iterations. The final device yields 27--29\% transmission at the top port and 67--68\% at the bottom port, achieving an insertion loss of approximately $0.16\,\text{dB}$ at $1,550\,\text{nm}$. In the case of the 10/90 power splitter, six wavelength sampling points were used, targeting 10\% transmission at the top port and 90\% at the bottom port. The optimization converges after 150 iterations, reducing the loss from 1.52 to $1.27 \times 10^{-3}$ (Fig.~\ref{fig:sfigure7}(c)). As illustrated in Fig.~\ref{fig:sfigure7}(d), the resulting device provides 7--8\% transmission at the top port and 85--87\% at the bottom port, corresponding to an insertion loss of about $0.26\,\text{dB}$. In all three optimization cases, convergence occurs around 150 iterations, indicating consistent optimization dynamics across the devices. This behavior is attributed to the identical latent space dimensionality used for all configurations. The final geometries of the optimized 50/50, 30/70, and 10/90 power splitters are displayed in Fig.~\ref{fig:sfigure7}(g)--Fig.~\ref{fig:sfigure7}(i), together with the normalized magnetic field distributions obtained from finite-difference frequency-domain (FDFD) simulations at 1,500, 1,550, and 1,600\,nm.

The transition from the EBL generator model to the PL generator model necessitates a strategic footprint expansion from $4.2 \times 4.2\,\mu\text{m}^2$ to $6 \times 6\,\mu\text{m}^2$. For the 50/50 power splitter, the EBL generator model achieves a loss of $3.2 \times 10^{-4}$ within 150 iterations, while the PL generator model converges to $2 \times 10^{-3}$ in the same number of iterations. Despite stricter constraints, the PL-based 50/50 power splitter maintains a 97\% overall transmission and a low insertion loss of $0.22\,\text{dB}$. This indicates that increasing the design area can partially offset the limitations imposed by a larger minimum feature size. The 30/70 power splitter further highlights the efficiency trade-offs between fabrication regimes. While both models converge within 150 iterations, the EBL generator model reaches a loss of $5 \times 10^{-4}$ with a $0.11\,\text{dB}$ insertion loss. Conversely, the PL generator model stabilizes at $1 \times 10^{-3}$ and exhibits a significantly higher insertion loss of $0.16\,\text{dB}$ at $1,550\,\text{nm}$. Despite this efficiency penalty, the device achieves the desired power distribution (27--29\% top port transmission, and bottom port transmission), proving the generator's ability to meet functional targets even when fine-grained topological features are restricted. In the high-contrast 10/90 power splitter case, optimization dynamics remain consistent, with both models reaching local minima after 150 iterations. The EBL generator model achieves a minimum loss of $4 \times 10^{-4}$ ($0.09\,\text{dB}$ insertion loss), outperforming the PL generator model, which stabilizes at $1.27 \times 10^{-3}$ with a $0.27\,\text{dB}$ insertion loss. The resulting 85--87\% bottom-port transmission for the PL design confirms the expected target operation. Ultimately, these small differences in performance are rooted in the reduced ``designable'' density of the parametrized manifold. The transition from an $18 \times 18$ latent space vector in the EBL generator model to a restricted $15 \times 15$ latent space vector in the PL generator model limits the available degrees of freedom. With fewer tunable parameters to satisfy electromagnetic boundary conditions, the optimization task becomes inherently more challenging. This necessitates the footprint expansion, accounts for the higher local minima, and explains the increased insertion loss observed in the PL generator model designs.

\subsection{Multi-Wavelength Photonic Duplexer}

A broadband wavelength duplexer was designed using the PL-based generator model to achieve efficient spectral routing within the $1,500$--$1,600\,\text{nm}$ wavelength range. The device directs optical signals between $1,500$ and $1,550\,\text{nm}$ to the upper output port and those between $1,550$ and $1,600\,\text{nm}$ to the lower port, thereby functioning as a short-pass and long-pass splitter, respectively. The total device footprint is $10.5 \times 10.5\,\mu\text{m}^2$. For optimization, we select ten discrete wavelengths within the $1,500$--$1,600\,\text{nm}$ range. Five wavelengths evenly distributed from $1,500$ to $1,550\,\text{nm}$ target 100\% transmission toward the top port, while the remaining five wavelengths between $1,550$ and $1,600\,\text{nm}$ target 100\% transmission to the bottom port. 

\begin{figure*}[ht!]
    \centering
    \includegraphics[width=\textwidth]{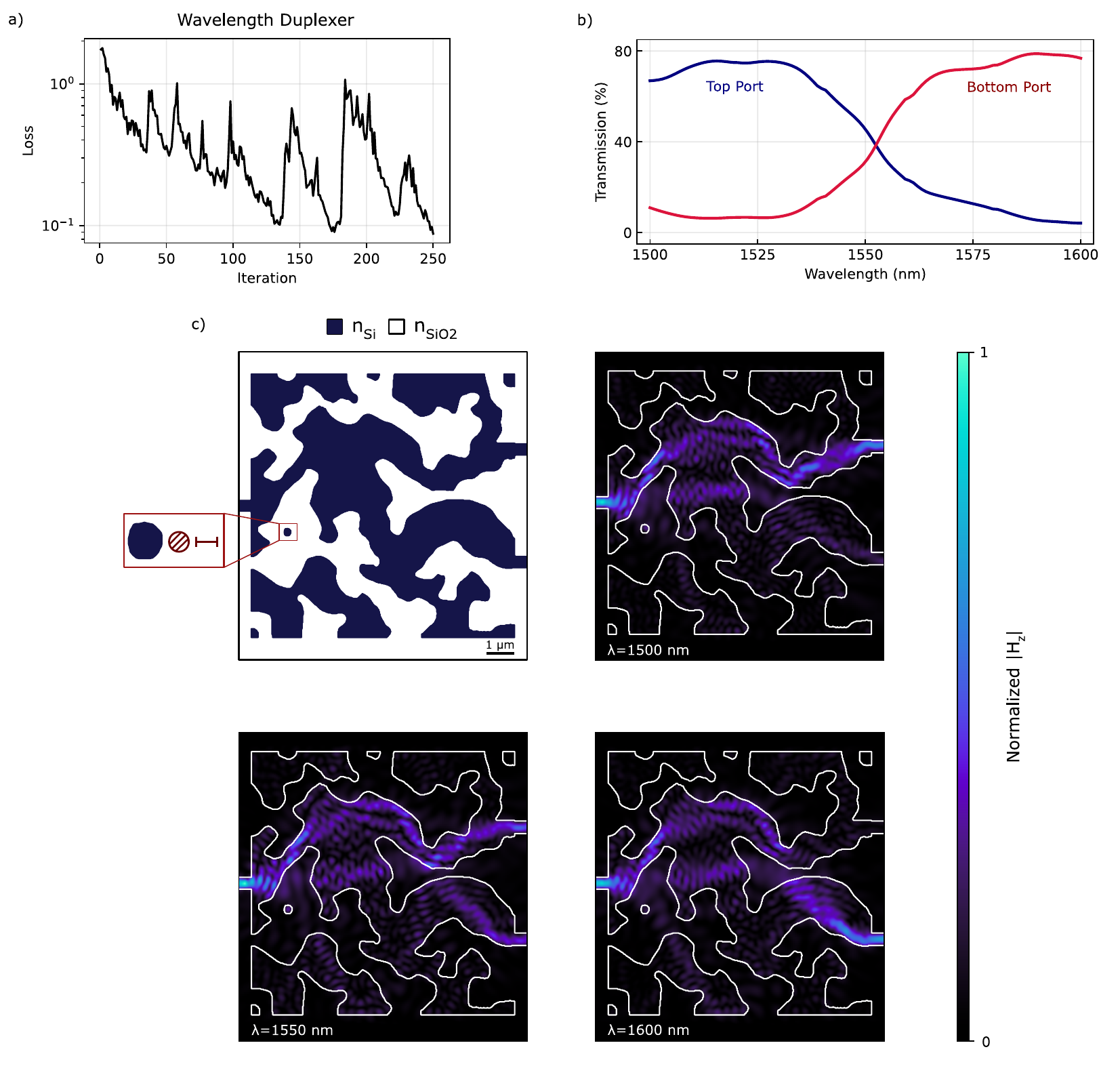}
    \caption{Multi-wavelength photonic duplexer optimized with a photolithography-compliant generator model. (a) Evolution of the optimization loss over 250 iterations, showing a significant drop from 1.74 to $8.7 \times 10^{-2}$. (b) Transmission spectra of the final device, illustrating effective separation between the top and bottom ports with a peak transmission of $\sim$94\%, corresponding to an insertion loss of 0.3~dB. (c) Final layout of the device (top left), with silicon regions in dark blue and SiO$_2$ in white, fully complying with design rule constraints. The adjacent panels display normalized FDFD magnetic field distributions at 1500, 1550, and 1600~nm, demonstrating consistent wavelength-dependent routing and confirming the duplexing performance.}
    \label{fig:sfigure8}
\end{figure*}

Following 250 optimization iterations, the loss value decreased substantially from 1.74 to $8.7 \times 10^{-2}$, as presented in Fig.~\ref{fig:sfigure8}(a). As shown in Fig.~\ref{fig:sfigure8}(b), the optimized device exhibits clear spectral separation between the two output channels, with an achieved transmission efficiency of approximately 94\%, corresponding to an insertion loss of $1.19\,\text{dB}$. The output waveguide separation for this photolithography-based wavelength duplexer is $4.5~\mu$m, measured center to center between the two output waveguides. The resulting geometry delineates the spatial distribution of $\text{Si}$ and $\text{SiO}_2$ regions and successfully passes DRC. Fig.~\ref{fig:sfigure8}(c) depicts the normalized FDFD magnetic field distributions at $1,500$, $1,550$, and $1,600\,\text{nm}$, illustrating the wavelength-dependent routing behavior and confirming the duplexing performance of the device.

The performance differences between the PL-based and EBL generator models highlight the critical trade-off between fabrication constraints and device optimality. While the PL model optimized through 250 iterations, it reached a minimum loss of only $8.7 \times 10^{-2}$. On the other hand, the EBL model achieved a superior loss of $1 \times 10^{-2}$ within 160 iterations. This performance gap is primarily attributed to the differences in the dimensionality of the latent space and the resulting feature size limitations. As the device exhibits a wavelength-dependent response, its performance is highly sensitive to subtle geometric variations that govern interference and phase accumulation across the structure. The PL model utilizes a $24 \times 24$ latent space grid, which provides fewer degrees of freedom compared to the $28 \times 28$ latent space employed by the EBL model. This reduced spatial resolution restricts the generator-based optimization capacity to encode fine geometric modulations necessary for shaping wavelength-dependent field distributions, leading to a degraded spectral response. This reduction in the optimized spatial degrees of freedom significantly decreases the tunability of the device geometry. Furthermore, the increased feature size (from $60\,\text{nm}$ to $150\,\text{nm}$) required for PL makes achieving the target performance more challenging, as the design space is more restricted. In contrast, the EBL--based generator model provides access to finer geometric degrees of freedom, enabling the optimizer to more accurately shape wavelength-dependent modal interference patterns and thereby achieve improved spectral performance. Consequently, the EBL model demonstrates enhanced transmission characteristics, reaching approximately 94\% efficiency compared to roughly 80\% observed in the PL design. This higher transmission translates directly to a lower insertion loss of $0.25\,\text{dB}$ for EBL, whereas the PL model is limited to $0.3\,\text{dB}$. Ultimately, while PL offers better scalability for manufacturing, the larger latent space and higher resolution of the EBL-based generator allow the optimizer to find a more efficient solution.

\subsection{Broadband Photonic Mode Converter}

A broadband mode converter was designed using the PL-based generator model, aiming to transform the fundamental $\text{TE}_0$ mode at the input into the first-order $\text{TE}_1$ mode at the output. The device occupies a compact footprint of only $2.75 \times 2.75\,\mu\text{m}^2$. During the optimization process, the target function was defined to maximize the modal overlap with the $\text{TE}_1$ electric field profile at the output port.

\begin{figure*}[ht!]
    \centering
    \includegraphics[width=0.85\textwidth]{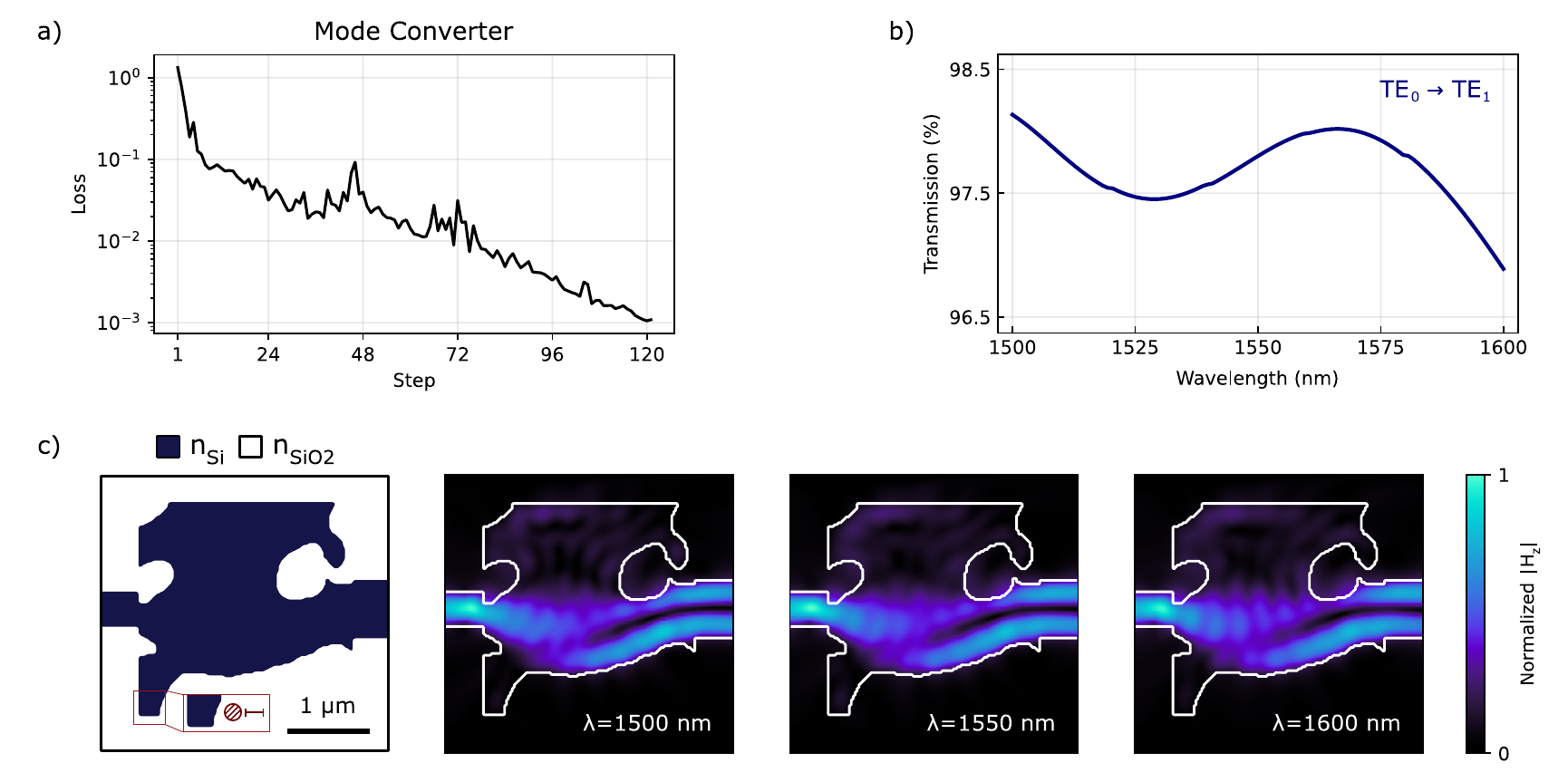}
    \caption{Broadband photonic mode converter designed with the photolithography-compliant generator model. (a) Evolution of the optimization loss over 120 iterations, decreasing from an initial 1.74 to a minimum of $9.5 \times 10^{-3}$. (b) Simulated transmission spectra showing a consistently high TE$_0$-to-TE$_1$ conversion ratio of 96--98\% across 1500--1600~nm, reaching $\sim$99\% efficiency at 1550~nm, corresponding to an insertion loss of only 0.03~dB. (c) Optimized permittivity profile of the device (left) alongside normalized magnetic field distributions ($|H_z|$) at wavelengths of 1500, 1550, and 1600~nm, highlighting successful broadband TE$_1$ mode conversion.}
    \label{fig:sfigure9}
\end{figure*}

As illustrated in Fig.~\ref{fig:sfigure9}(a), the optimization converged after approximately 120 iterations, with the loss decreasing from an initial value of 1.74 to a minimum of $9.5 \times 10^{-3}$. The transmission into the $\text{TE}_1$ mode remained consistently high throughout the optimization, fluctuating between 96\% and 98\% (Fig.~\ref{fig:sfigure9}(b)). The final transmission efficiency of $\sim$99\% corresponds to an insertion loss as low as $0.1\,\text{dB}$. The optimized device geometry is depicted in Fig.~\ref{fig:sfigure9}(c). The input waveguide width is $450\,\text{nm}$, gradually expanding to $700\,\text{nm}$ at the output. Field distributions simulated at $1.50$, $1.55$, and $1,600\,\text{nm}$ confirm the broadband behavior of the device. The normalized FDFD field maps clearly demonstrate the successful conversion into the $\text{TE}_1$ mode, characterized by the emergence of two symmetric intensity lobes at the output waveguide.

The observed performance disparity, where the EBL generator model converges to a loss of $3 \times 10^{-4}$ while the PL generator model reaches $9.5 \times 10^{-3}$, is again a direct consequence of the trade-off between fabrication constraints and design freedom. Specifically, as the minimum feature size increases from $60\,\text{nm}$ to $150\,\text{nm}$, the PL generative model is forced to generate larger-scale features, which inherently lack the fine-grained control to manipulate ultrabroadband light. Consequently, while the EBL generator-based designs achieve near-ideal performance within $2.1 \times 2.1\,\mu\text{m}^2$ footprint, the PL generator-based design requires an expanded $2.75 \times 2.75\,\mu\text{m}^2$ area to redistribute the modal power and compensate for the reduced efficiency of larger dielectric features. Furthermore, the optimization for the PL generator-based device design is conducted in a lower-dimensional $8 \times 8$ latent space vector, compared to the $10 \times 10$ latent space vector used for the EBL generator-based design, which imposes a more rigid constraint on the available parameter space. This reduction in tunable degrees of freedom necessitates a more exhaustive search for an optimal solution, explaining the increased iteration count required to navigate the high-dimensional cost landscape under stricter geometric priors. Despite these fabrication limitations, both devices exhibit exceptional performance: the PL generator model achieves a transmission efficiency of approximately 99\% ($0.03\,\text{dB}$ insertion loss), while the EBL-based model achieves a transmission efficiency of approximately 98\% ($0.1\,\text{dB}$ insertion loss). These results confirm that our generative approach can successfully achieve complex modal transformations using fabrication-compatible geometries across different lithographic regimes.

\section{Symmetric 50/50 Power Splitter Design}

The applications requiring a near-exact 50:50 splitting ratio and a 0-degree 
phase offset between the outputs benefit from symmetric device geometries. Our 
design framework can in fact enforce such symmetry without requiring full 
retraining of the generative model. To achieve this, the latent space is 
partitioned along the length-wise axis of the power splitter, effectively 
treating only half of the design region as the independent optimization 
variables. Following each forward pass, the resulting binary permittivity map is 
mirrored across this symmetry axis to produce a fully symmetric design. A similar 
morphological closing operation is subsequently applied to prevent the formation 
of enclosed voids near the ``half-device'' boundaries, ensuring that the final 
geometry remains both compliant with design rules. This procedure integrates with 
our existing latent-space optimization pipeline without any overhead, and requires 
no modification to the trained generator models.

\begin{figure*}[ht!]
    \centering
    \includegraphics[width=0.8\textwidth]{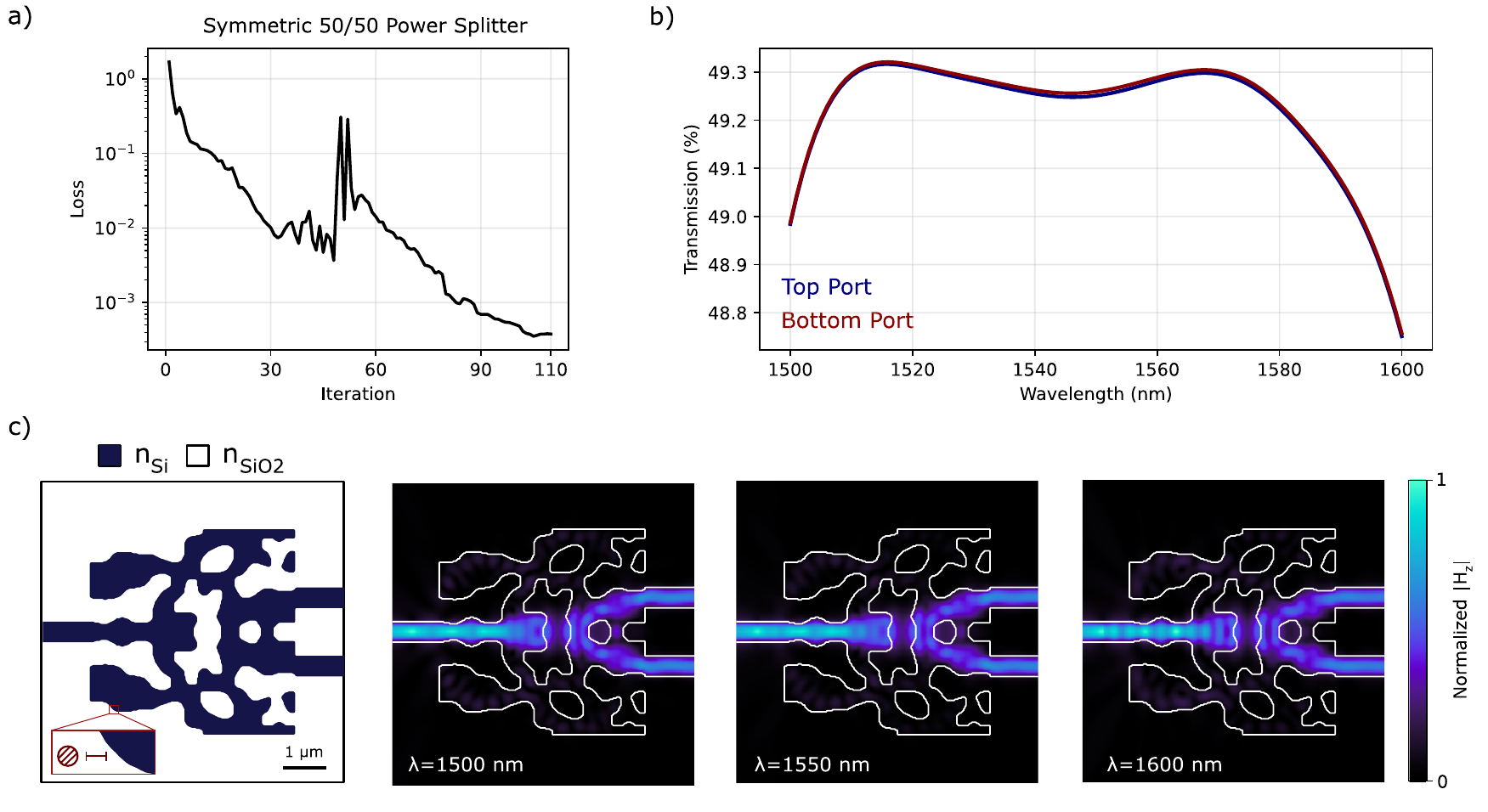}
    \caption{Design of a symmetric 50:50 power splitter with electron-beam-lithography 
generator model. (a) Optimization progress showing convergence to a mean squared 
error of $2.0 \times 10^{-4}$ within 110 iterations. (b) Transmission spectra 
for top and bottom ports across the 1500--1600~nm wavelength range. The design 
achieves a low insertion loss of 0.067~dB at 1550~nm with inherently in-phase 
outputs. (c) Optimized device geometry generated by partitioning the latent space 
to enforce structural symmetry. Optimized device geometry and normalized magnetic 
field magnitudes $|H_z|$ at $\lambda = 1500$~nm, $\lambda = 1550$~nm and 
$\lambda = 1600$~nm.}
    \label{fig:sfigure10}
\end{figure*}

We validated this approach by designing an EBL-compatible, symmetric 50:50 power 
splitter. The results shown in Fig.~\ref{fig:sfigure10} demonstrate that the method reaches 
convergence after 110 iterations. At an MSE of ${\sim}10^{-4}$ both the 
symmetry-enforced ($2.0 \times 10^{-4}$) and original non-symmetric 
($2.8 \times 10^{-4}$) devices converge to similar quality results within 
run-to-run stochastic variance of the optimizer. Moreover, enforcing symmetry 
also reduces the effective latent dimensionality by half. The reduced iteration 
count (compared to 149 iterations for the non-symmetric counterpart) reflects 
this lower-dimensional search induced by the symmetry constraint. Across the full 
wavelength range from 1500 to 1600~nm, the device maintains a near-perfect 
simulated 50:50 power distribution between the two output ports. The resulting 
insertion loss is 0.067~dB at 1550~nm. Furthermore, the device is fully 
symmetric, resulting in in-phase outputs by construction. These results confirm 
that application-specific symmetry configurations can be enforced within our 
generative design framework through structural constraints, without sacrificing 
device performance.

\section{Fabrication Tolerance Across All Designed Devices}

To analyze the performance variation with fabrication, we have performed a 
systematic tolerance analysis for all designed devices by simulating $\pm 15$~nm 
their uniformly over-etched and under-etched versions. The pixel-based 50/50 
power splitter exhibits slightly better tolerance than its generator-based EBL 
counterpart. This can be attributed to the different constraint regimes of the 
two methods. Pixel-based optimization enforces design constraints gradually 
through the filtering and scheduled projection operations. Naturally, this 
optimization trajectory passes through grayscale intermediate states (as shown 
in Fig.~3d) that effectively average over small geometric perturbations. In 
contrast, the generator enforces DRC compliance strictly throughout optimization, 
leaving less room for the emergence of tolerance-favoring device geometries. 
Nevertheless, we note that the manifold formulation can be easily extended to 
incorporate fabrication tolerance during optimization. Typical workflows for this 
include multi-objective optimizations where as-designed, over-etched, and 
under-etched versions of the same device are optimized simultaneously 
\cite{WangFengwen2022}. Alternatively, the generator training objective can 
also be augmented with a tolerance-aware regularization \cite{LazarovBoyan2016}. These 
approaches can enable fabrication tolerance in the existing generator framework 
without retraining the generator models.

Across both generators, all devices maintain target functionality with some 
shifts in the transmission spectra under $\pm 15$~nm etching perturbations. 
Comparative results indicate that PL-optimized devices tend to outperform their 
EBL counterparts in fabrication tolerance. This is expected since a $\pm 15$~nm 
perturbation represents a smaller fractional deviation relative to the 150~nm PL 
minimum feature size than to the 60~nm EBL features. The transmission deviations 
and spectral shifts are consistent with the fabrication-induced changes typically 
reported for inverse-designed silicon photonic devices 
\cite{AlexanderPiggott2015, PiggottAlexander2017, PiggottAlexander2020, Molesky2018, LoganSu2018, WangFengwen2022, ChenYuchen2023}.

\begin{figure*}[ht!]
    \centering
    \includegraphics[width=\textwidth]{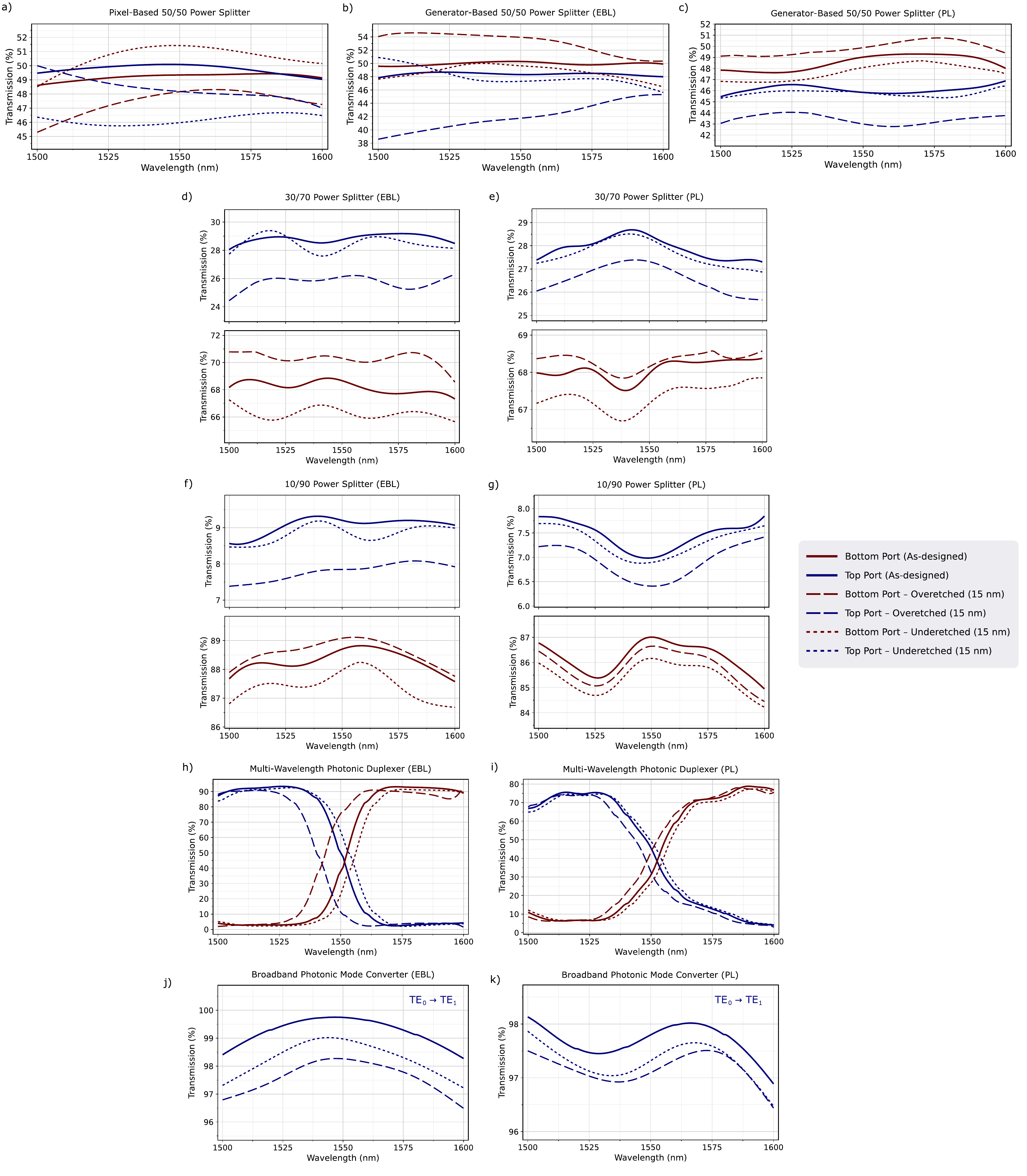}
    \caption{Fabrication tolerance analysis of all designed devices under $\pm 15$~nm etch 
perturbations. Transmission spectra are shown for the as-designed geometry (dark 
solid lines), over-etched (dashed lines), and under-etched (dotted lines) 
structures, for the top (blue) and bottom (red) output ports. (a) Pixel-based 
50/50 power splitter; (b, c) generator-based 50/50 power splitters for EBL and 
PL models; (d, e) 30/70 and (f, g) 10/90 power splitters, (h, i) 
multi-wavelength photonic duplexers and (j, k) broadband photonic mode 
converters.}
    \label{fig:sfigure11}
\end{figure*}

\newpage
\section{Pixel-Based vs. Generator-Based Design for Wavelength Duplexer on 150 nm Photolithography Platform}

\begin{figure*}[ht!]
    \centering
    \includegraphics[width=0.7\textwidth]{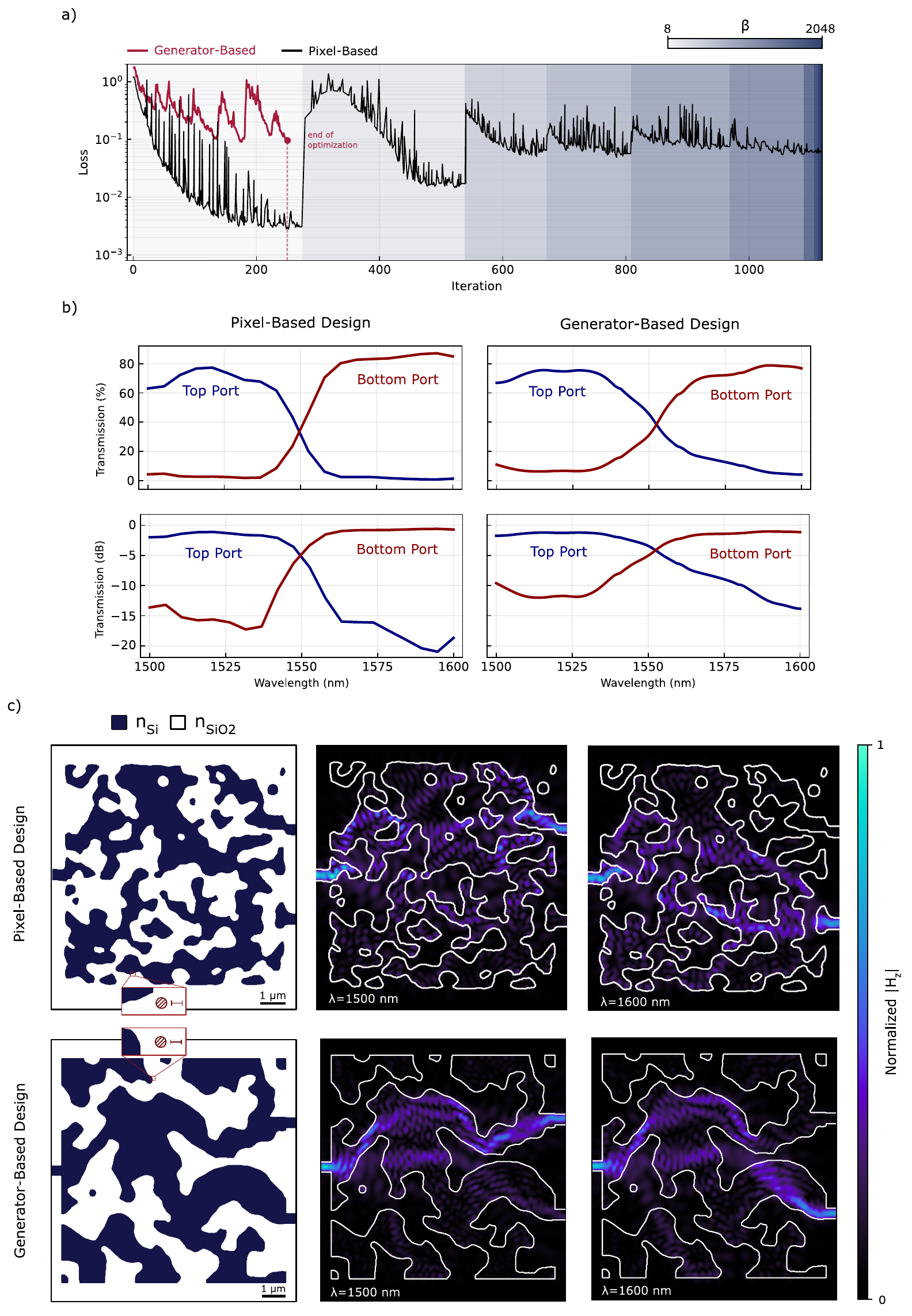}
    \caption{Pixel-based vs.\ generator-based inverse design of a wavelength duplexer on the 
150~nm photolithography platform. (a) Optimization loss as a function of 
iteration for the two methods. The generator-based method converges in 250 
iterations, while the pixel-based method requires 1,116 for convergence. (b) Linear-scale transmission spectra (\%) and corresponding dB-scale transmission spectra
for the two output ports of the duplexers. (c) Device 
geometries and normalized magnetic field magnitudes $|H_z|$ at $\lambda = 1500$~nm 
and $\lambda = 1600$~nm.}
    \label{fig:sfigure12}
\end{figure*}

\newpage

To further validate the generality of our approach, we extend the comparison 
between pixel-based and generator-based inverse design to the wavelength duplexer 
geometry on the more fabrication-constrained 150~nm photolithography platform. 
The 150~nm PL platform provides reduced access to sub-wavelength features 
relative to 60~nm EBL. and the wavelength duplexer is a more constrained design 
problem than the 50/50 splitter. These are both consistent facts with the 
intuition that the set of DRC-compliant optimal duplexer topologies is smaller. 
For this reason, we have optimized the duplexer on a larger 
$10.5 \times 10.5~\mu$m$^2$ footprint. The results shown in Fig.~\ref{fig:sfigure12} are 
consistent with what we observed for the 50/50 splitter in Fig.~3. The 
generator-based method converges in 250 iterations to a final loss of 
$8.7 \times 10^{-2}$, while the pixel-based method requires 1,116 iterations and 
achieves a loss of $5.4 \times 10^{-2}$. This indicates a 4.5-times difference 
in convergence speed, similar with the results of the 50/50 splitter. As before, 
the slowest $\beta$ update schedule and the L-BFGS optimizer was used for the 
pixel based results, to reach the lowest losses possible in the pixel based 
approach (see Supplementary Information Section~10 for comparative details). The 
transmission plots for the two approaches show that both methods produce 
functional wavelength duplexers on the 150~nm photolithography platform. The 
generator-based approach reaches a comparable solution in a fraction of the 
iterations, while the pixel-based approach achieves a marginally better final 
loss at substantially higher computational cost.

\section{Convergence Behavior and Optimizer Dependence of Pixel-based Device Design}

The increases of the loss function observed in Fig.~3a reflect the algorithmic 
behavior of the optimizer itself. The way a given optimizer traverses the loss 
landscape according to its algorithmic configuration can be significantly 
different from another optimizer; and this may yield significantly different 
optimization trajectories. Two distinct types of oscillations are present in the 
loss curve, and are addressed separately below. The way a given optimizer 
traverses the loss landscape according to its algorithmic configuration can be 
significantly different from another optimizer; and this may yield significantly 
different optimization trajectories. To illustrate these details, we have added a 
new Supplementary Information section covering the algorithmic origin of these 
features in the loss curve, the convergence behavior itself, and a comparison for 
the identical 50/50 splitter across three well-known optimizers. The large upward 
jumps at discrete points in the loss curve reflect the $\beta$ continuation 
strategy used to drive the design toward a fully binary structure (consisting of 
Si and SiO$_2$ only). Each time $\beta$ is doubled, the objective landscape 
changes discontinuously, and the optimizer must adapt to the new structure of 
this loss surface. These doublings are indicated by the background color change 
in Fig.~3a, and also in Fig.~\ref{fig:sfigure13} below. This is the standard accepted 
strategy for obtaining fabricable binary designs in topology optimization 
\cite{WangFengwen2022, LazarovBoyan2016}. These ``continuation strategy transitions'' are 
therefore an inherent feature of the schedule used. The rapid oscillations within 
a single $\beta$ stage in the L-BFGS optimized result in Fig.~3(a) have a 
separate reason. At each iteration, L-BFGS proposes a candidate step, evaluates 
the objective, and rejects the step if certain conditions are not satisfied, and 
backtracks to a smaller step size \cite{wolfe1969, liu1989}. Since 
every function evaluation (including these rejected proposals) is recorded, each 
rejected step appears as an upward spike followed by a return to a lower value. 
This is the expected signature of line search in L-BFGS. The loss at the end of 
each $\beta$ stage is lower than at the end of the previous stage. This 
convergence is consistent with a correctly computed objective and a corresponding 
gradient.

\begin{figure*}[ht!]
    \centering
    \includegraphics[width=\textwidth]{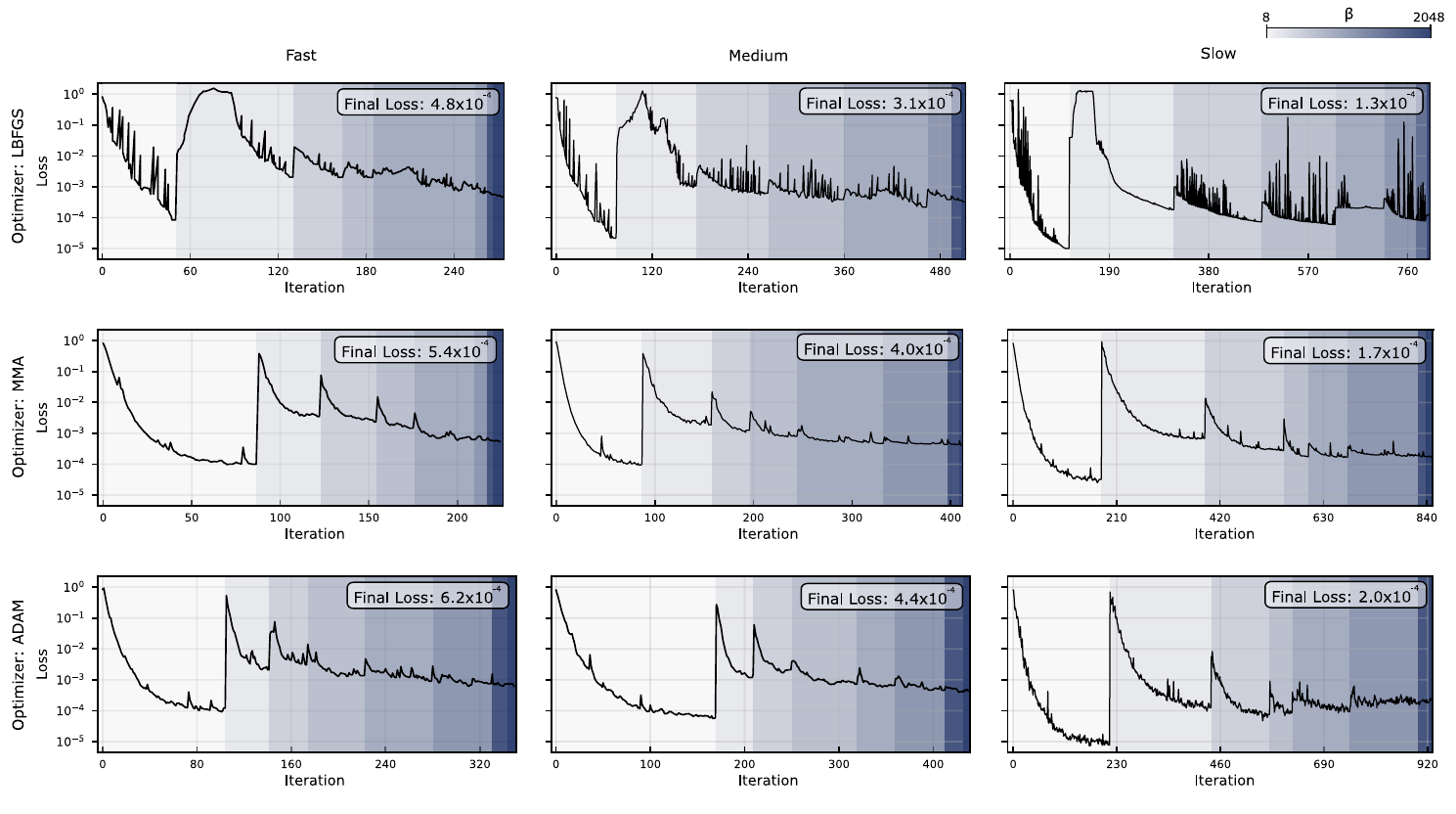}
    \caption{Loss curves for the pixel-based projection optimization across three optimizers 
(rows: L-BFGS, MMA, ADAM) and three $\beta$ scheduling strategies (columns: 
fast, medium, slow). Stage-transition jumps ($\beta$ doublings) appear in all 
three optimizers. Within-stage oscillations are characteristic of L-BFGS 
line-search backtracking, and smaller occasional bumps in MMA and ADAM reflect 
their own step-size dynamics.}
    \label{fig:sfigure13}
\end{figure*}

\newpage

To further characterize the optimizer-dependence of these trajectories, we test 
convergence using different optimizers as well as using faster or slower $\beta$ 
scheduling strategies. The results of this study are shown in Fig.~S13. Here, 
we test three optimizers commonly used in photonic design tasks (L-BFGS 
\cite{liu1989}, MMA \cite{svanberg1987}, and ADAM \cite{kingma2014}). For each 
one, we also test fast, medium, and slow doubling strategies for the $\beta$ 
parameter by checking relative convergence of loss during the last 2, 3, and 5 
iterations, respectively. The results show that for a given $\beta$ update 
schedule, all three optimizers converge to comparable final losses. The 
trajectories themselves reveal that L-BFGS experiences the most significant 
spikes throughout the optimization, with some also observed for MMA and ADAM. 
These can be attributed to the optimizers' step-size dynamics where MMA's moving 
asymptotes can overshoot under poor local approximations \cite{svanberg1987}, and 
ADAM's momentum and adaptive learning rate can produce transient bumps if 
gradients change sharply \cite{reddi2018}. This indicates once again that 
oscillations in the loss function are inherent to the specific optimization 
routine chosen. As expected, the best optimization results (lowest final loss 
values) were achieved using the slow update schedule across all three optimizers. 
In the slow update schedule, L-BFGS also consistently achieved the lowest final 
loss in the fewest total iterations. Therefore, L-BFGS under slow $\beta$ 
scheduling was chosen as the reference for the main text in Fig.~3a, because 
it represents the best-performing pixel-based configuration out of all nine 
combinations, with the lowest final loss of $1.3 \times 10^{-4}$.

\newpage

\def\bibsection{\section*{References}}
\bibliographystyle{naturemag}
\bibliography{supplement}